\documentclass[onecolumn,showpacs,preprintnumbers,amsmath,amssymb]{revtex4-1}

\usepackage{amsfonts}
\usepackage{amsmath}
\usepackage{amssymb}
\usepackage{graphicx}
\usepackage{dcolumn}
\usepackage{bm}
\usepackage{tabularx}
\usepackage{epstopdf}

\begin{document}


\title{Early-warning signals of topological collapse in interbank networks}

\author{Tiziano Squartini}
\affiliation{Instituut-Lorentz for Theoretical Physics, Leiden Institute of Physics, University of Leiden, Niels Bohrweg 2, 2333 CA Leiden, The Netherlands}
\author{Iman van Lelyveld}
\affiliation{De Nederlandsche Bank, PO box 98, 1000 AB Amsterdam, The Netherlands}
\author{Diego Garlaschelli}
\affiliation{Instituut-Lorentz for Theoretical Physics, Leiden Institute of Physics, University of Leiden, Niels Bohrweg 2, 2333 CA Leiden, The Netherlands }


\date{\today}

\begin{abstract}
The financial crisis clearly illustrated the importance of characterizing the level of `systemic' risk associated with an entire credit network, rather than with single institutions. However, the interplay between financial distress and topological changes is still poorly understood. Here we analyze the quarterly interbank exposures among Dutch banks over the period 1998-2008, ending with the crisis. After controlling for the link density, many topological properties display an abrupt change in 2008, providing a clear -- but unpredictable -- signature of the crisis. By contrast, if the heterogeneity of banks' connectivity is controlled for, the same properties show a gradual transition to the crisis, starting in 2005 and preceded by an even earlier period during which anomalous debt loops could have led to the underestimation of counter-party risk. These early-warning signals are undetectable if the network is reconstructed from partial bank-specific data, as routinely done. We discuss important implications for bank regulatory policies.
\end{abstract}

\pacs{Valid PACS appear here}
\maketitle

\section*{Introduction}

Financial and banking systems are strongly interconnected networks of institutions exposed to both endogenous and exogenous fluctuations \cite{may0,may1}.
When defaults occur, they cascade throughout the network and can cause the collapse of an entire system, as dramatically witnessed by the recent financial crisis \cite{subprime}.
As a consequence, the analysis of economic and financial networks as the propagation channel for distress has received a lot of attention \cite{science,newman2010,goyal2009,jackson,iman,iman2,iman3,iman4,bargall,bargall2,simon,myPRE1,myPRE2,mynull,mattmars}.
Much effort has been devoted to the search for regularities in the structure of financial networks, i.e. looking for degree heterogeneity, a core-periphery or a modular structure \cite{iman,iman2,iman3,iman4,bargall}.
Similarly, null models have been introduced in order to understand whether part of the observed topological complexity can be explained relatively simply in terms of the observed heterogeneity of vertices \cite{bargall2,simon,myPRE1,myPRE2,mynull}.
For interbank networks specifically, a lot of attention has been devoted to quantifying the level of \emph{systemic risk} (the risk of the collapse of the system as a whole) determined by a particular network topology, as opposed to the traditional measures of risk defined for indidividual banks \cite{mattmars,simulations,debtrank,eisenberg}.
It turns out that the minimization of (standard measures of) individual risk can often increase the level of systemic risk, which in turn can hurt individual financial entities \cite{may0,may1}. This highlights the inadequacy of  traditional models and regulation and suggests an analogy with ecological networks \cite{myfoodwebs,foodwebmotifs}.

All the above approaches focus on the structural properties (e.g. systemic risk), associated with a given, static network topology.
However, interbank networks are highly dynamic. 
As distress starts to propagate, existing financial connections might dissolve and new ones might materialize, modifying the way further reverberations of a crisis are channeled through the network.
Therefore, 
one needs to take into account both the (expected) effects of network topology on the stability of the financial system and the reverse effects of (realized) defaults on the structure of interbank networks.
This has led to models of interbank networks that dynamically adapt to critical events, with a continuous feedback between topology and dynamics \cite{co-pierre}.

In this paper, 
rather than introducing a theoretical model, we carry out an empirical characterization of the interplay between realized financial stress and the changes in the observed interbank structure. We address two main questions: \emph{does the topology of an interbank network undergo major structural change as a crisis suddenly manifests itself?} And if so, \emph{are there any topological precursors of this structural change, to be used as early-warning signals of the approaching crisis?}
Our results indicate that the answer to both questions is affirmative. 

\section*{Results}

\subsection*{Data}

 For our analysis we focus on the recent global financial crisis, that manifested itself at the end of 2007 and continued throughout 2008 \cite{subprime}, and on its build-up phase, which is much more difficult to identify.
We selected a dataset reporting 44 quarterly snapshots of the Dutch Interbank Network (DIN in the following), starting from the first quarter of 1998 and ending with the last quarter of 2008 \cite{iman2}.
Each snapshot reports the exposures between Dutch banks at the end of the corresponding quarter, and represents them as binary links directed from the borrower to the lender.
Our data include the year when the crisis manifested itself in its strongest form (2008) plus the preceding 10 years, arguably the build-up phase. More details about the data are given in the Supplementary Information (Appendix).
Note that, when studying the propagation of defaults, the binary topology of interbank networks plays the primary role. The magnitude of the connections, while surely important, plays mainly a quantitative role.
For instance, the existence and uniqueness of a `clearing payment vector' that clears the obligations of all banks after a default only depends on the interbank topology \cite{eisenberg}.
Moreover, while a weighted network is of course more informative than its binary projection, recent empirical results \cite{myPRE1,myPRE2,mynull} have shown that the knowledge of a binary property often conveys more information about a real-world economic network than the knowledge of the corresponding weighted property. 

\subsection*{Topological signatures of the crisis}

\begin{figure}[t!]
\centering
\includegraphics[width=0.99\textwidth]{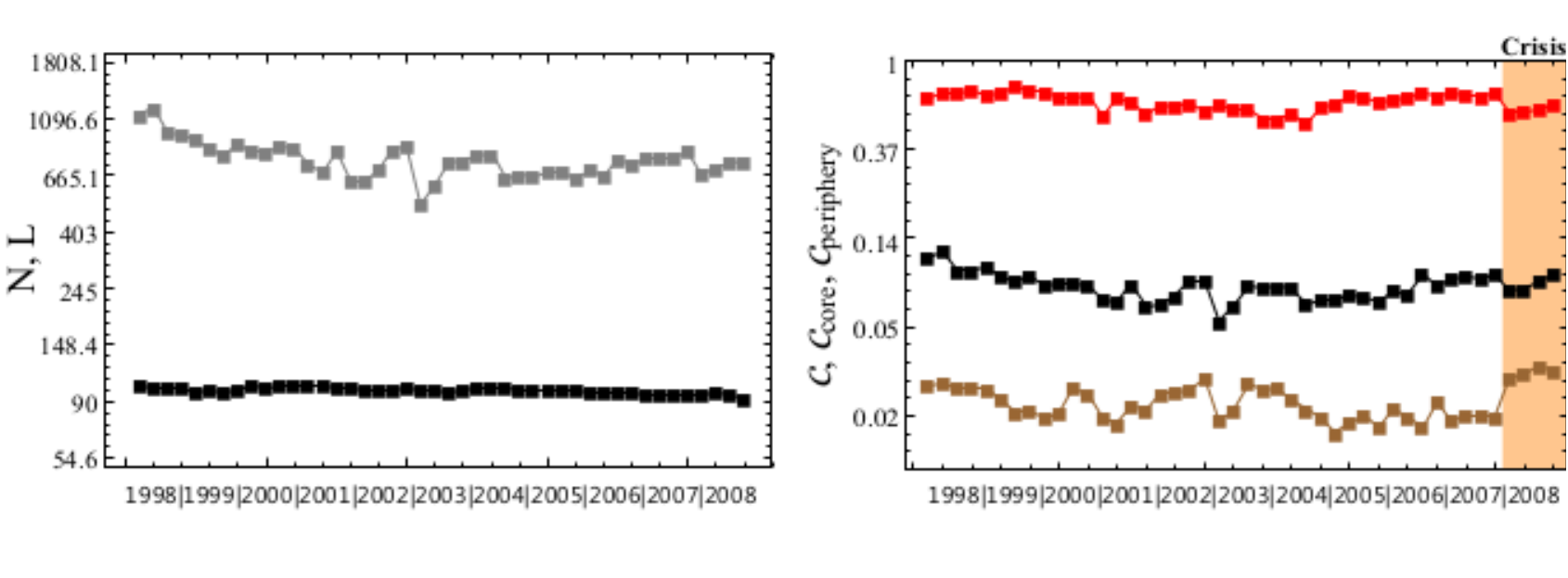}
\caption{Left: observed number of vertices (black) and links (gray). Right: observed density of the whole network (black), of the core (red) and of the periphery (brown). The $y$-scale is logarithmically spaced in both cases.
}
\label{conn}
\end{figure}

We start by looking for possible topological signatures of the crisis.
We find that, at the onset of the crisis the size (numbers of vertices $N$), the connectedness (number of links $L$) and the  link density (or connectance $c$) of the network (see fig.~\ref{conn} right) do not show any significant change in their (roughly stationary) trends (see fig.~\ref{conn} left).

We can also separately consider the density of the core and of the periphery of the network \cite{iman2}: the ideal core-periphery model (CP model) assumes that core banks are all bilaterally linked with each other, that periphery banks do not lend to each other, and that core banks both lend to and borrow from at least one periphery bank (discussed further in \cite{iman2} and Appendix).
From the right panel of fig.~\ref{conn}, we confirm that the core is much denser than the periphery (as by construction should be).
However, the core- and periphery-specific densities, exactly as the overall density, show only a slight jump from the end of 2007 to the beginning of 2008. The size of the change is not significant, as it is of the same order as the fluctuations characterizing the entire 11-years time interval.
Taken together, the above results show that the size and density of the network (as well as their
core-specific and periphery-specific values) are completely uninformative about the crisis. 

\begin{figure}[t!]
\hspace{3.5cm} {\bf DRG} \hspace{6.5cm} {\bf DCM}
\centering
\includegraphics[width=.99\textwidth]{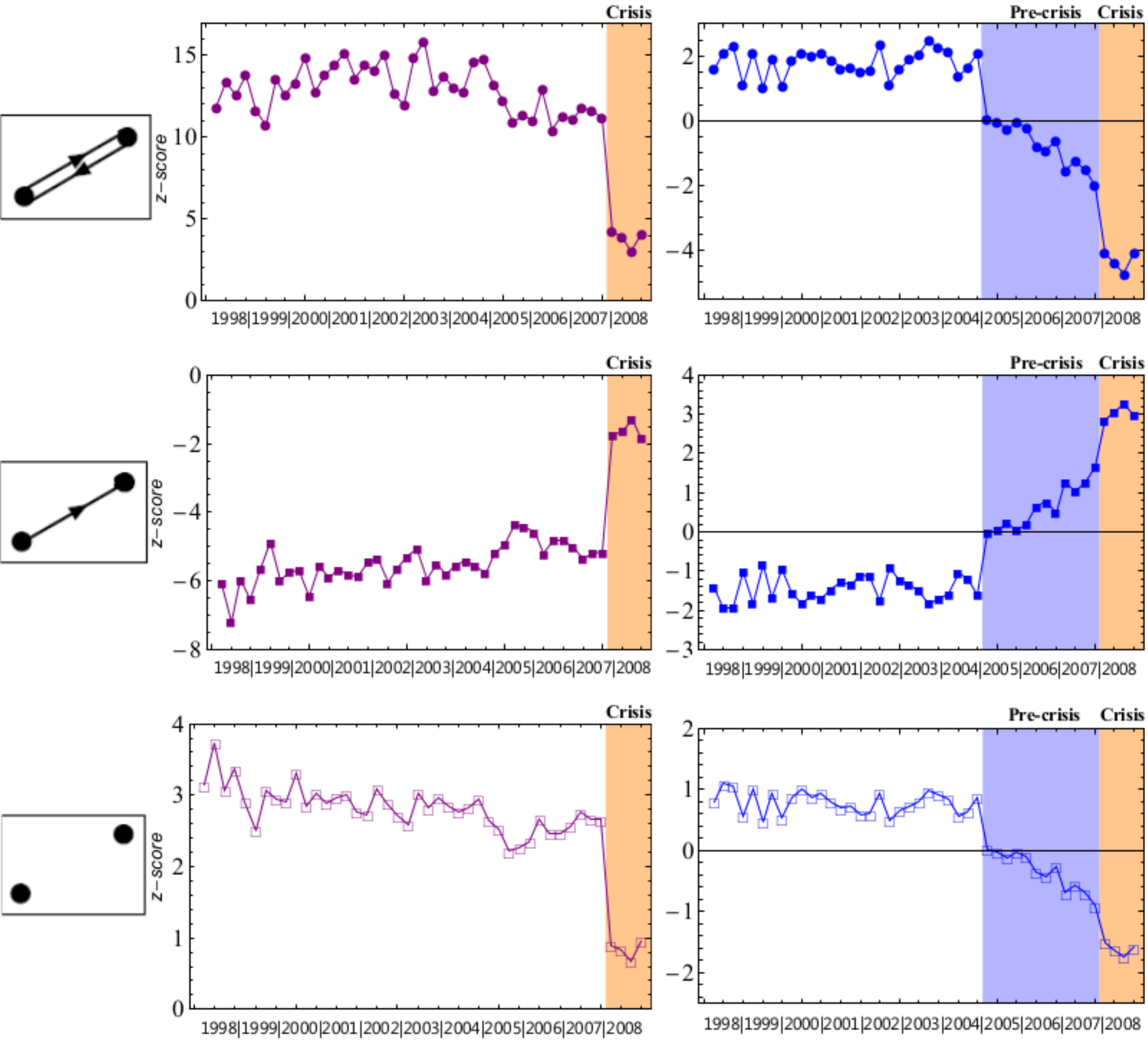}
\caption{Temporal evolution of the dyadic $z$-scores: $z_{L^{\leftrightarrow}}$ under the DRG (top-left, purple circles) and the DCM (top-right, blue circles), $z_{L^{\rightarrow}}$ under the DRG (middle-left, purple, full squares) and the DCM (middle-right, blue, full squares), $z_{L^{\nleftrightarrow}}$ under the DRG (bottom-left, purple, empty squares) and the DCM (bottom-right, blue, empty squares).}
\label{dydrg}
\end{figure}

However, we are going to show that the picture changes if, after \emph{controlling} for the size and density themselves, we consider higher-order topological properties (dyadic, triadic, and so on).
We first focus on the relative frequency or abundances of the three possible \emph{dyadic motifs} in the observed network, i.e. the number $L^{\leftrightarrow}$ of reciprocated (full) dyads, the number $L^{\rightarrow}$ of non-reciprocated (single) dyads, and the number $L^{\nleftrightarrow}$ of empty dyads (see fig.~\ref{dydrg}).
These numbers are informative only after filtering out size and density effects, or even more complicated topological properties.
Therefore, here and in what follows, we compare each measured quantity $X$ with the expected value $\langle X\rangle$ under a \emph{null model} which has some properties in common with the observed network but is otherwise maximally random. More precisely, we introduce z-scores (see Methods section) to quantify the deviation between data and null model.
Technically, the method we adopt is an analytical and unbiased one \cite{mymethod} based on maximum-entropy ensembles of graphs with constraints \cite{newman_expo} (see Appendix for details). We stress that the use of a null model is very different from that of a proper explanatory model: throughout the entire paper, we do not aim at introducing a model that accurately reproduces the data. Rather, the null  models we define represent different benchmarks, with various levels of complexity, that discount for the immediate effects of  certain topological properties treated as constraints. Comparing the data with the predictions of a null model allows us to determine which  observed structural properties are not simply explained by the constraint specifying the null model itself. Indeed, our most informative findings will correspond to a \emph{deviation}, rather than an  \emph{agreement}, with null models. It should therefore be clear that null models are by construction in-sample, as it would make no sense to control, in one snapshot of the network, for the effects of a topological property observed in a different snapshot. The inherently in-sample nature of null models is very different from the out-of-sample one of explanatory models, where the fit with one snapshot of the data is used to reproduce different snapshots.

In the left panels of fig.~\ref{dydrg} we show the evolution of the $z$-scores over time \cite{foodwebmotifs,motifs,motifs2,mymotifs} for each of the three dyadic motifs, under a null model that controls for the size and density of the network. The \emph{Directed Random Graph} (DRG), the directed version of the Erd\H{o}s-R\'enyi random graph model (see Appendix) is such null model. Note that the $z$-score is a convenient measure as it immediately shows which motifs diverge significantly from the null model (see Methods section).
We find that, while the size and density of the network are relatively stable throughout the entire period, all the dyadic $z$-scores show an abrupt jump in 2008.
The crisis period (highlighted in ochre in fig.~\ref{dydrg}) is characterized by a sudden decrease of the abundance of full and empty dyads, and a sudden increase of the abundance of single dyads.
Note that, before the crisis, the number of reciprocated dyads is significantly larger than the expected one, while during the crisis it becomes marginally consistent with the null model (i.e., the random graph).

Similarly, the observed abundances of single and empty dyads become consistent with the null model in 2008.
Since the total number of links is more or less stable, the net effect we see is that reciprocal connections suddenly `decouple' and fill previously empty dyads, making single dyads increase and empty dyads decrease with respect to their expected abundance level.
So the network seems to suddenly evolve from a fluctuating but roughly stationary configuration (with few single dyads and many full and empty ones) to a `crisis' configuration whose dyadic structure is marginally consistent with that of an unstructured random graph.
We denote this sudden loss of structure as the `collapse' of the original network.

The dyadic motifs, when using the DRG as a reference, are therefore clear topological signatures of the crisis.
They are however not predictive, since they show an abrupt transition with no evidence of a preceding build-up phase.
They allow us to `see' the crisis, but not to `foresee' it.

\subsection*{Early-warning topological precursors: the pre-crisis phase}

Surprisingly, the picture changes completely if we consider a more stringent null model where the intrinsic heterogeneity of banks is accurately controlled for.
In particular, we compare each snapshot of the network with a null model (known as the \emph{Directed Configuration Model} (DCM)) where the number of in- and out-going links of each bank is kept equal to the observed values. For a discussion of the properties of the degree distribution(s) of this network, see \cite{iman2}.

Note that the DRG is an unlikely benchmark economically as the degrees of all banks are narrowly distributed around their empirical average value and banks are thus forced to be similar in size. As a consequence, when studying the deviations of the real network from the DRG, we cannot disentangle the effects of unrealistic bank-specific properties from those of genuine higher-order (dyadic and beyond) patterns.
Incidentally, this shows the main limitation of the \emph{representative agent} concept when applied to economic networks \cite{representative}.
By contrast, the DCM indirectly preserves the real heterogeneity of banks, by preserving the observed degrees produced by that heterogeneity.
This provides a realistic benchmark with deviations indicating a genuine signature of higher-order effects beyond the bank-specific level, directly arising from the choices of banks.

The second column of fig.~\ref{dydrg} shows the dyadic $z$-scores under the DCM.
When comparing these values with the previous ones obtained under the DRG, we find surprising results.
Firstly, during the first seven years (quarters 1-28, i.e. from 1998 to 2004 included) all $z$-scores are stationary and have the same sign as under the DRG, but are much closer to zero. Their small absolute value ($|z|\lesssim 2.5$) suggests that during this period the dyadic structure of the network is not far from the prediction obtained under the DCM, i.e. it is roughly explained by the heterogeneous degrees of banks.
By contrast, starting from the 29th quarter (2005Q1), all $z$-scores suddenly change sign and start to move away from their previous stationary values.
This gradually leads to the collapsed network configuration of 2008. The network is then the most distant from the DCM (and, as we observed before, the closest to the DRG).
However, the `collapse' is not a sudden structural change, as it is clearly preceded by a 3-year `pre-crisis' period (from 2005 to 2007 included, highlighted in purple in fig.~\ref{dydrg}) bridging the earlier stationary phase to the 2008 crisis.

Our results shown so far suggest that, compared to a homogeneous benchmark, the interbank network displays an abrupt structural transition at the onset of the crisis. In contrast, with a heterogeneous benchmark, the transition is slow and continuous and highlights a gradual build-up phase starting three years in advance of the crisis. The pre-crisis phase is thus an early-warning signal of the upcoming topological collapse.

Topological patterns captured by dyadic motifs are limited to correlations within pairs of vertices.
In order to study a higher level of organization, we now analyse the \emph{triadic motifs} \cite{foodwebmotifs,motifs,motifs2,mymotifs}, i.e. the possible patterns involving three connected vertices.

\begin{figure}[t!]
\centering
\includegraphics[width=.89\textwidth]{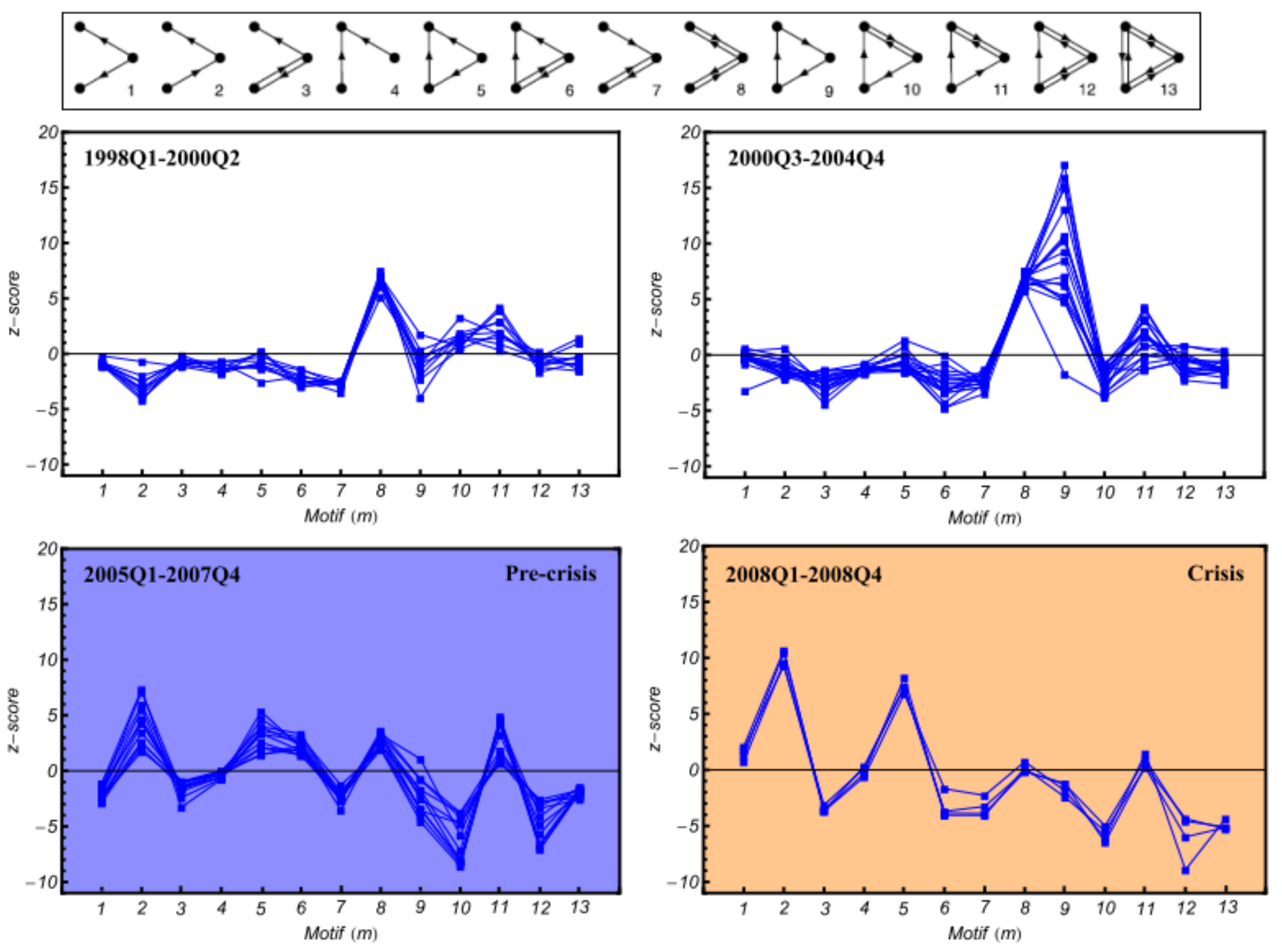}
\caption{Triadic $z$-scores for all the 44 quarters, grouped into four subperiods, under the DCM. First subperiod: $t_{1}$-$t_{10}$ (top-left); second subperiod: $t_{11}$-$t_{28}$ (top-right); third subperiod: $t_{29}$-$t_{40}$ (bottom-left); fourth subperiod: $t_{41}$-$t_{44}$ (bottom-right).}
\label{mot_1}
\end{figure}

Exactly as for the dyads, we consider the $z$-scores for the abundances of each of the 13 triadic motifs (see Appendix for definitions).
However, before considering the development of individual $z$-scores over time, we first identify the most significant motifs by comparing all 13 $z$-scores with each other in sub-periods.
This results in the `motif profiles' \cite{motifs,mymotifs} shown in fig.~\ref{mot_1}, where we used the DCM as the null model.

It turns out that the 44 quarterly snapshots do not collapse to a single profile (see Appendix).
By contrast, we can clearly distinguished four subperiods with different characteristic profiles, as evident from the four panels of fig.~\ref{mot_1}. Remarkably, we find that the last two subperiods coincide exactly with the pre-crisis (2005-2007) and crisis (2008) periods we identified before.

\begin{figure}[t!]
\hspace{2.8cm} {\bf DRG} \hspace{5.2cm} {\bf DCM}
\centering
\includegraphics[width=.80\textwidth]{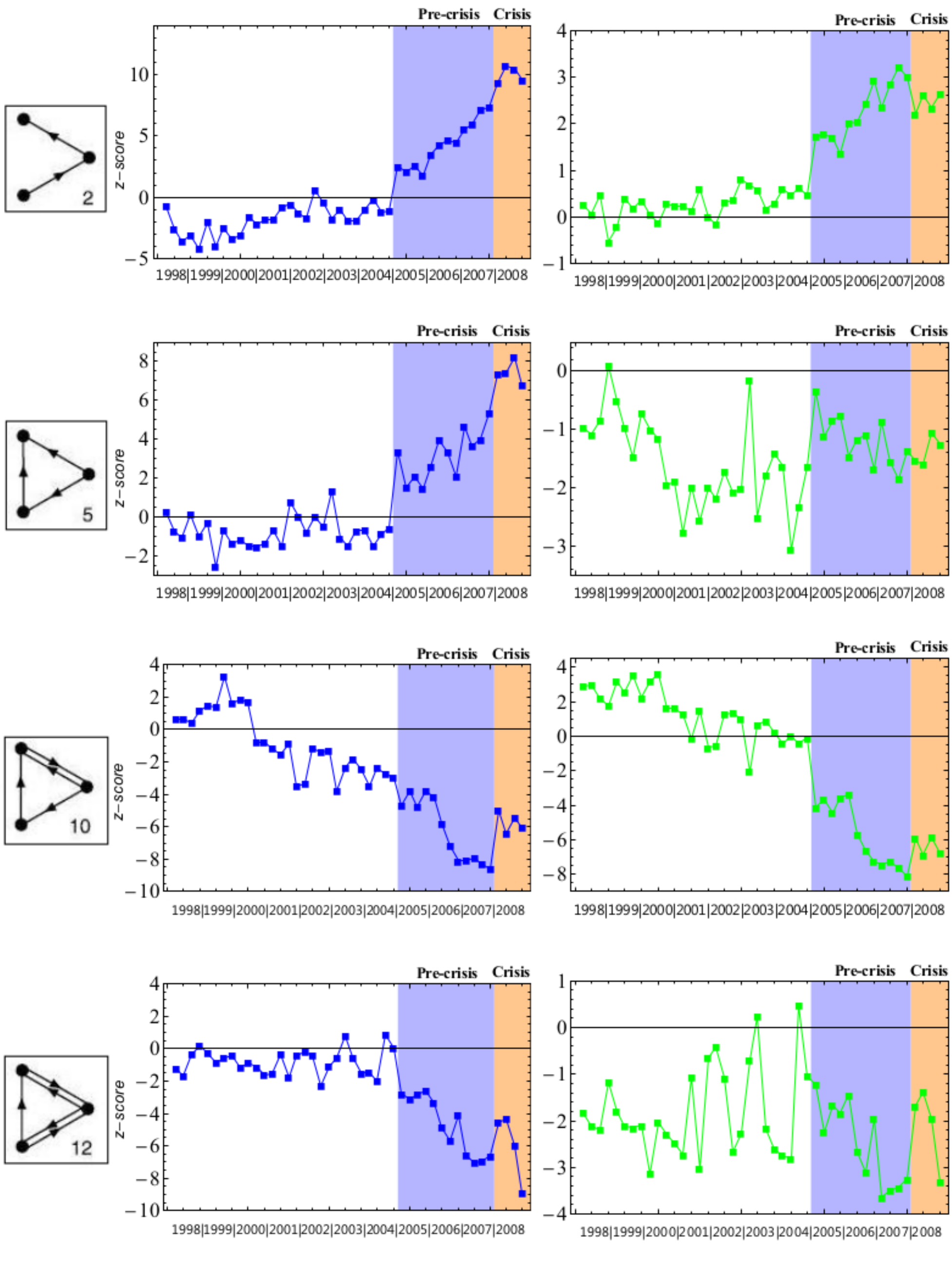}
\caption{Temporal evolution of the triadic $z$-score for motifs 2, 5, 10 and 12 under the DCM (left) and the RCM (right).}
\label{mot_4t}
\end{figure}

As is clear from the bottom right panel of fig.~\ref{mot_1}, we can identify the motifs number 2, 5, 10 and 12 as the most significant ($|z|\gtrsim 4.5$) triadic signatures of the 2008 crisis.
If we now track these motifs over time (see left panels of fig.~\ref{mot_4t}), we find exactly the same behavior as shown above for the dyads: the trends over the entire 1998-2004 period are stationary (with small $z$-scores indicating an approximate accordance with the DCM), and from 2005 onwards they gradually evolve towards the collapsed configuration (for motif 10 the departure actually starts before 2005, but this anomaly will be corrected by a more constrained null model, as we show below). This confirms that the building-up of the crisis is undetectable under homogeneous assumptions, while it becomes manifest in the gradual divergence of the real interbank market from the configuration expected on the basis of the observed heterogeneity of banks.

Clearly, since triads are combinations of dyads, some triadic motifs might be over(under)-represented just because the dyadic motifs they contain are over(under)-represented, in which case the triad as a whole should not be considered an interesting pattern \emph{per se}. 
In order to control for this, we introduce a more stringent null model that separately controls for the number of single and reciprocated links of each vertex (\emph{Reciprocal Configuration Model} (RCM), see Appendix for details). The RCM separately preserves the number of empty, full, and single (out- and inward) dyads in which each vertex is involved.
As a result, all the observed dyadic abundances are preserved and the dyadic $z$-scores are zero by construction.
In the right panels of fig.~\ref{mot_4t} we show the triadic $z$-scores for the same motifs considered previously, but now recalculated under the RCM.
We find that motif 2 shows the same trend as before and motif 10, falling in line with it, now confirms the beginning of the pre-crisis phase in 2005.
This indicates that motifs 2 and 10 are important building blocks of the network.
Motifs 5 and 12 are instead no longer significant ($|z|\lesssim 3.5$), and their fluctuating trends do not show any appreciable change during the pre-crisis and crisis periods.

\subsection*{The earliest precursor: anomalous circular lending}

What remains to be explained is the nature of the separation (occurring in the mid of 2000) between the first two subperiods shown previously in fig.~\ref{mot_1}, as all the trends considered so far do not display any significant change at that particular point in time.

\begin{figure}[t!]
\centering
\includegraphics[width=.89\textwidth]{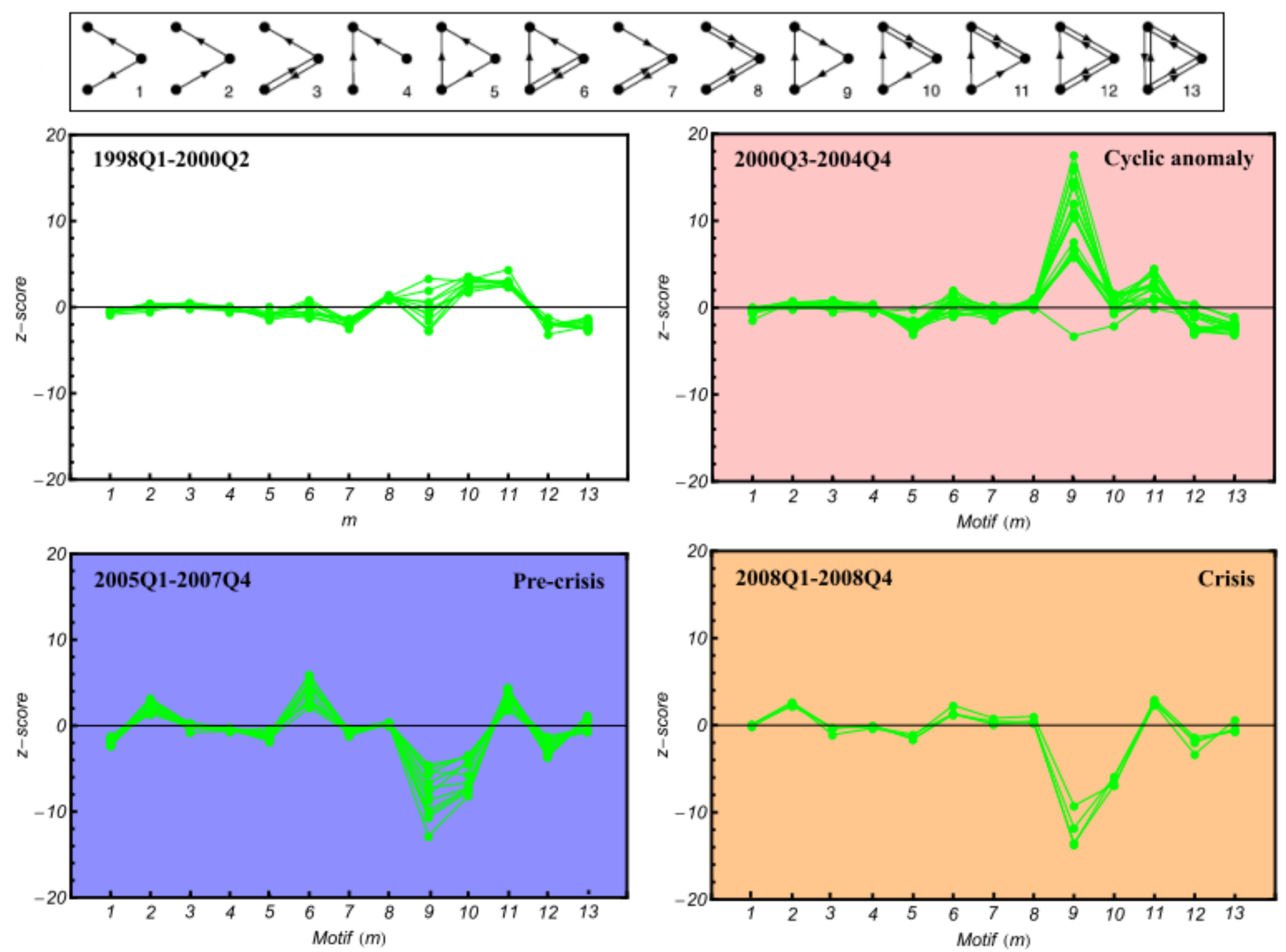}
\caption{Triadic $z$-scores for all the 44 quarters, grouped into four subperiods, under the RCM. First subperiod: $t_{1}$-$t_{10}$ (top-left); second subperiod: $t_{11}$-$t_{28}$ (top-right); third subperiod: $t_{29}$-$t_{40}$ (bottom-left); fourth subperiod: $t_{41}$-$t_{44}$ (bottom-right).}
\label{mot_12}
\end{figure}

Before answering this question, we stress that although fig.~\ref{mot_1} might look quite different under the RCM, we find similar results in that case as well.
As before, the motif profiles calculated under the RCM over the entire 1998-2008 period do not collapse to a universal distribution (see Appendix).
Still, inside each of the four subperiods we identified earlier, the profiles are coherent (see fig.~\ref{mot_12}).

Remarkably, the first regime is now almost completely consistent with the null model ($|z|\lesssim 4$ for all 13 motifs), which means that the heterogeneous local connectivity and reciprocity of banks entirely explain the triadic structure.
Moreover, if we now look more closely at figs.~\ref{mot_1} and \ref{mot_12}, we find that the main differences between the first two subperiods are determined by motifs 9 (under both the DCM and RCM) and 10 (under the DCM, but not the RCM).

Thus the only significant change occurring in the middle of 2000, after controlling for the dyadic structure, is due to motif 9. The temporal evolution of the latter is reported in fig.~\ref{mot_9}, under both null models.
We find that, from the third quarter of 2000 to the last quarter of 2004, motif 9 indeed shows a marked difference with respect to the rest of the period, and turns out to be strongly over-represented, highlighting an anomalously high number of triads of banks involved in circular lending loops with no reciprocation.
Since this subperiod is only characterized by the over-representation of motif 9 (all other motifs are still approximately consistent with the RCM), we denote it as the `cyclic anomaly' phase (highlighted in pink in figs.~\ref{mot_12} and \ref{mot_9}), and regard it as the earliest precursor of the 2008 crisis.
Remarkably, when the cyclic anomaly phase ends and the pre-crisis phase begins, motif 9 suddenly changes from being the strongest over-represented to being the strongest under-represented motif under the RCM (while not significant under the DCM).
Thus, it appears that non-reciprocated lending loops, that were the arrangement preferred by triads of banks before 2005, suddenly became the `most avoided' triad.
The following two periods (pre-crisis and crisis) are indeed mainly characterized by an increasingly strong under-representation of motifs 9 and 10 (see figs.~\ref{mot_4t}, \ref{mot_12} and \ref{mot_9}), which both involve a circular lending loop.

\section*{Discussion}

The above results have potentially strong implications for bank regulation policies. An immediate one is that the popular view that real interbank markets consist of a well defined core-periphery structure, and consequently that banks can be binarily classified either as big/central or as small/peripheral, is far too simple. Our findings show that the observed heterogeneity of banks is irreducible to the core-periphery dichotomy. Rather, the opposite is true: given the observed heterogeneity of banks, the network is found to have no significant core-periphery structure, and sometimes even has an `anti-core' one (see Appendix).

The approximate consistency between the real network and the RCM in the initial 1998-2000 period also suggests that, in absence of distress, the topology of real interbank networks might be quite accurately reconstructed using only the knowledge of the number of (inward, outward, and reciprocated) partners of each bank.
Technically, this means that, under low stress, real interbank networks might be typical members of an equilibrium statistical ensemble of graphs, where banks' connectivities are maximally informative. In practical terms, it means that to characterize the network, data requirements are very limited.

However, and more importantly, our findings also show that during the build-up of crises the network can keep moving away from the expectations derived only from the knowledge of bank-specific properties. 
In this out-of-equilibrium regime, the local connectivities of banks become less and less informative about the network as a whole.
This loss of topological predictability speaks against the use of maximum-entropy techniques aimed at reconstructing the most likely configuration of an (unobserved) interbank network when only local information about the total assets and liabilities of each bank is available \cite{simon}.
Since assets and liabilities are the (transaction) weighted counterparts of the in- and out-degree of a bank, our results suggest that this technique might yield a realistic guess of the real network only in tranquil times.
When the network is under stress, maximum-entropy would instead provide a greatly distorted picture of it.

Strikingly, if our analysis had been carried out on the most likely network consistent with the observed degrees (i.e. the DCM), then every dyadic, triadic, or core-periphery property would have appeared, at each point in time, as perfectly consistent with the configuration model. Supervision based only on bank-specific information, and not on the knowledge of the entire network, is thus likely to remain oblivious to warning signals of structural changes in the run-up to the crisis.

Moreover, both reciprocity and triadic structure have implications for counterparty risk assessment.
As other authors have already pointed out \cite{co-pierre,acharya}, Over-the-Counter (OTC, i.e. not disclosed to third parties) transactions intrinsically generate \emph{risk externalities}: if bank A issues loans to banks B and C, it will require an interest rate that depends on the estimated counterparty risk (which is a function of the fluctuating financial variables on which the `health' of banks B and C depend).
But if B issues another loan to C, and A is not aware of this, the interest rate charged by A will underprice counterparty risk, since B becomes vulnerable to a default of C, increasing the correlation between the health of B and that of C.
Note that this particular triad is motif number~5.
This example shows that the binary topology of interbank networks (more than the intensity of links) has direct effects on systemic risk, and also highlights that some triadic motifs are strongly affected by risk externalities.

Now, it should be noted that the unreciprocated 3-loop (motif number 9) maximizes the underestimation of risk in OTC transactions: each of the three banks involved is not aware of the fact that counterparty risk loops back to itself, creating correlations not incorporated in their bilateral risk pricing.
This suggests that, during the cyclic anomaly phase, banks might have systematically underestimated risk externalities.

\begin{figure}[t!]
\centering
\includegraphics[width=.99\textwidth]{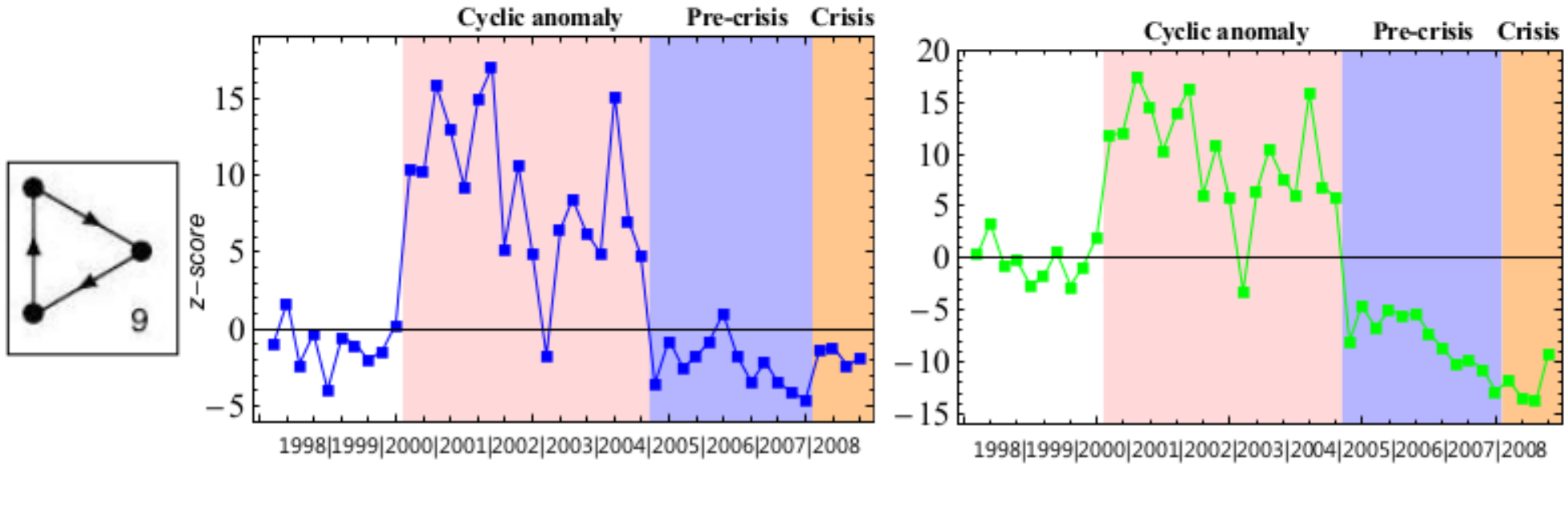}
\caption{Temporal evolution of the triadic $z$-score for motif number 9, under the DCM (left) and under the RCM (right).}
\label{mot_9}
\end{figure}

Note that circularity itself is not necessarily associated with strong risk externalities; but \emph{unreciprocated} circularity is.
For instance, within a full dyad, risk loops back between the two banks as well. But in this case both parties are aware of it, and can properly include the increased correlation in their risk pricing.
Also note that, while full dyads are still prone to the risk externality involving a third party, this will be a smaller effect since the probability that risk loops back along a longer chain of defaults is smaller than that of risk looping back within the dyad itself.
Thus, at a dyadic level, \emph{single} dyads are the most prone to the underestimation of counterparty risk, precisely because they can become parts of unreciprocated loops.
Again, this effect is purely topological: in a mutual dyad with two positive but strongly asymmetric weights, both banks can still incorporate these weights to properly price their risk.
Only if one weight is zero, i.e. in a single dyad, this is no longer possible. This further explains why the key information relevant to us is encoded in the binary topology, and not in the intensity of connections.
By contrast, at a triadic level, 3-loops involving an increasing number of reciprocated dyads (motifs number 9, 10, 12 and 13 respectively, see Appendix) are increasingly less prone to the risk externality.
Unreciprocated loops (of any length greater than 2) can therefore be considered to be a sort of `autocatalytic risk loops'.
Since longer loops imply smaller probabilities of cascading defaults, presumably the most dangerous autocatalytic risk loops are precisely those involving three banks.

During the cyclic anomaly phase, all the partly reciprocated loops (motifs 10, 12, 13) were much less abundant than the completely unreciprocated 3-loop (and always consistent with both null models (DCM and RCM)) thus increasing systemic risk.
During the pre-crisis phase, the loops with small or no reciprocation (motifs 10 and 9) became increasingly under-represented (figs. \ref{mot_4t} and \ref{mot_9}).
Unfortunately, during the same period reciprocated dyads (that dominated the earlier phase) also became increasingly under-represented, and outnumbered by single dyads (fig.~\ref{dydrg}).
This suggests that, starting from 2005, the underestimation of systemic risk might have progressively increased, first due to autocatalytic risk loops during the cyclic anomaly phase and on a simpler, dyadic level during the pre-crisis phase.

These considerations show that OTC transactions have the potential to create unintentional but emergent, self-reinforcing and destabilizing patterns and feeds into the debate on how OTC markets can be monitored and regulated. Since our results on `risk autocatalysis' suggests that, even when banks spontaneously engage in reciprocated transactions, autocatalytic risk loops can emerge (in the cyclic anomaly phase the reciprocity is still high, cf fig~\ref{mot_12}), simply requiring that banks reciprocate a fair amount of transactions (unless this means \emph{all} transactions) is not enough to prevent the creation of unreciprocated loops.
One possibility is to introduce a Central Clearing Counterparty (CCP) who would step in the middle of bilateral OTC trades.
Although this would reduce the systemic risk due to private interaction, it does introduce the (systemic) risk that the CCP can fail as well.
Another approach is to start properly monitoring OTC markets, acting on anomalous motifs. Given the data collection efforts underway for instance in the UK and internationally at the Bank for International Settlements this could be concretely considered (\cite{Caruana,Langfield}).

Although our results are strong on providing early warning signals, they cannot explain the economic rationale for the observed network patterns. As the links are formed in an OTC market, the participants only knowingly create the dyads, not the triadic motifs. A standard explanation is that financial markets are used to hedge risks \cite{Freixas}. Unexpected idiosyncratic shocks are covered in the market, either through a cash transaction or through a longer running (derivative) transaction. If things change, which is likely given that shocks arrive continuously, banks generally do not close out a contract but take out a second contract in the opposite direction. Over a reporting period this would lead to significant gross exposure (but much lower net exposure). As during the Great Moderation prior to the crisis, shocks were small, the need to enter into ever more insurance/hedging contracts diminished and thus the probability that any uni-directional link would turn into a reciprocal link became smaller.

However, since the `phase' transition is taking place both in tranquil and stressed times we need a second explanation for our result: in tranquil times everyone became an acceptable counterparty so hedges no longer needed to be effectuated with existing, known counterparties. Therefore reciprocal dyads slowly became under represented. As the crisis arrived, however, banks seem to actively try to find unconnected parties and thus the existing development towards fewer reciprocal dyads intensified. Such behaviour might be driven by bank's
aiming for more diversification, both on the asset and the liability side, achieved by breaking existing links and forming links with hitherto unrelated nodes.

In sum, our results clearly indicate that further theoretical and empirical research is needed to understand the economics of network formation. More generally, any policy directed at regulating interbank markets in a `pairwise' fashion appears to be fundamentally ineffective, since the most significant patterns are found to occur at an irreducibly triadic level. This result moves the regulation target even further away: while the notion of systemic risk already implies that monitoring individual banks is insufficient to contain systemic risk, monitoring pairs of banks is also likely to fail; the minimal `building blocks' appear to be triples of banks.

\section*{Methods}

In order to detect the statistically significant deviations of a measured quantity
$X$ from the expected value $\langle X\rangle$, we calculated standard deviation $\sigma[X]$ under the null model and define the $z$-score
\begin{equation}
z_{X}\equiv\frac{X-\langle X\rangle}{\sigma[X]}
\label{eq:z}
\end{equation}
The $z$-score is a standardized variable measuring the difference between the observed and the expected value in units of standard deviation.
If $X$ is normally distributed under the null model, then values within $z=\pm1,\:z=\pm2,\:z=\pm3$ would (approximately) occur with a $68\%,\:95\%,\:99\%$ probability respectively.
If the observed value of $X$ corresponds to a large positive (negative) value of $z_{X}$ then the quantity $X$ is over(under)-represented in the data, and not explained by the null model.
For most of the topological properties we consider (i.e. dyads and core-periphery structure), the normality under the null model is either trivially ensured by the Central Limit Theorem (CLT), or checked numerically (see Appendix).
The CLT cannot be invoked for triads due to statistical dependencies among the random variables involved (triads necessarily share dyads, and are therefore not independent of each other). Still, larger z-scores identify more significant patterns.
Even if null models do not represent, by themselves, a forecasting procedure, it is nevertheless possible to use our analysis to detect a temporal trend, once the deviations highlighted by the z-scores analysis are plotted versus time. The resulting trends, clearly underlining an ongoing structural change, can be interpreted as the starting point of a predictive inference procedure.

\section*{appendix\label{app}}

\section{Data\label{sectwo}}

Data are taken from the database used in \cite{iman2} by using prudential reporting of balance sheet positions of Dutch banks. The data includes all the exposures between Dutch banks (from contractual obligations to swaps) up to one year and of more than 1.5 million euros, on a quarterly reported frequency (from 1998Q1 to 2008Q4) \cite{iman2}. In other words, the data covers forty-four time-periods, corresponding to the forty-four ends of quarters of eleven years (from 1998 to 2008) and shows only the existence (or not) of exposures between (anonymized) Dutch banks of more than 1.5 million euros.
Note that the last four temporal snapshots correspond to the year 2008, i.e. the fir st year of the self-evident crisis, whose beginning can traced back to August 2007 (as already pointed out in \cite{iman2}), i.e. the sixth-last temporal snapshot considered here.

Given the nature of the available data, the Dutch interbank network (DIN) is represented as a \emph{binary}, \emph{directed network} where vertices represent banks and links represent exposures: a link pointing from bank $i$ to bank $j$, at time $t$, indicates the existence of (at least) an exposure of more than 1.5 million euros, directed from $i$ to $j$, registered at the end of the particular  quarter $t$.
The number $N$ of banks varies from period to period, oscillating between 91 and 102 (see main text).
For each quarter $t$ (with $t=1,\dots, 44$), the structure of the network is therefore entirely described by an $N\times N$ (in general asymmetric) adjacency matrix $A$, whose entries are $a_{ij}=1$ if a binary directed link from bank $i$ to bank $j$ exists during that quarter, and $a_{ij}=0$ otherwise.
The number $L$ of directed links is therefore computed as
\begin{equation}
L\equiv \sum_{i=1}^N\sum_{j(\ne i)=1}^N a_{ij}
\end{equation}
And the link density, or connectance $c$, is
\begin{equation}
c\equiv\frac{L}{N(N-1)}
\end{equation}

\section{The Core-Periphery Model\label{somsec:cp}}

\begin{figure}[t]
\centering
\includegraphics[width=.4\textwidth]{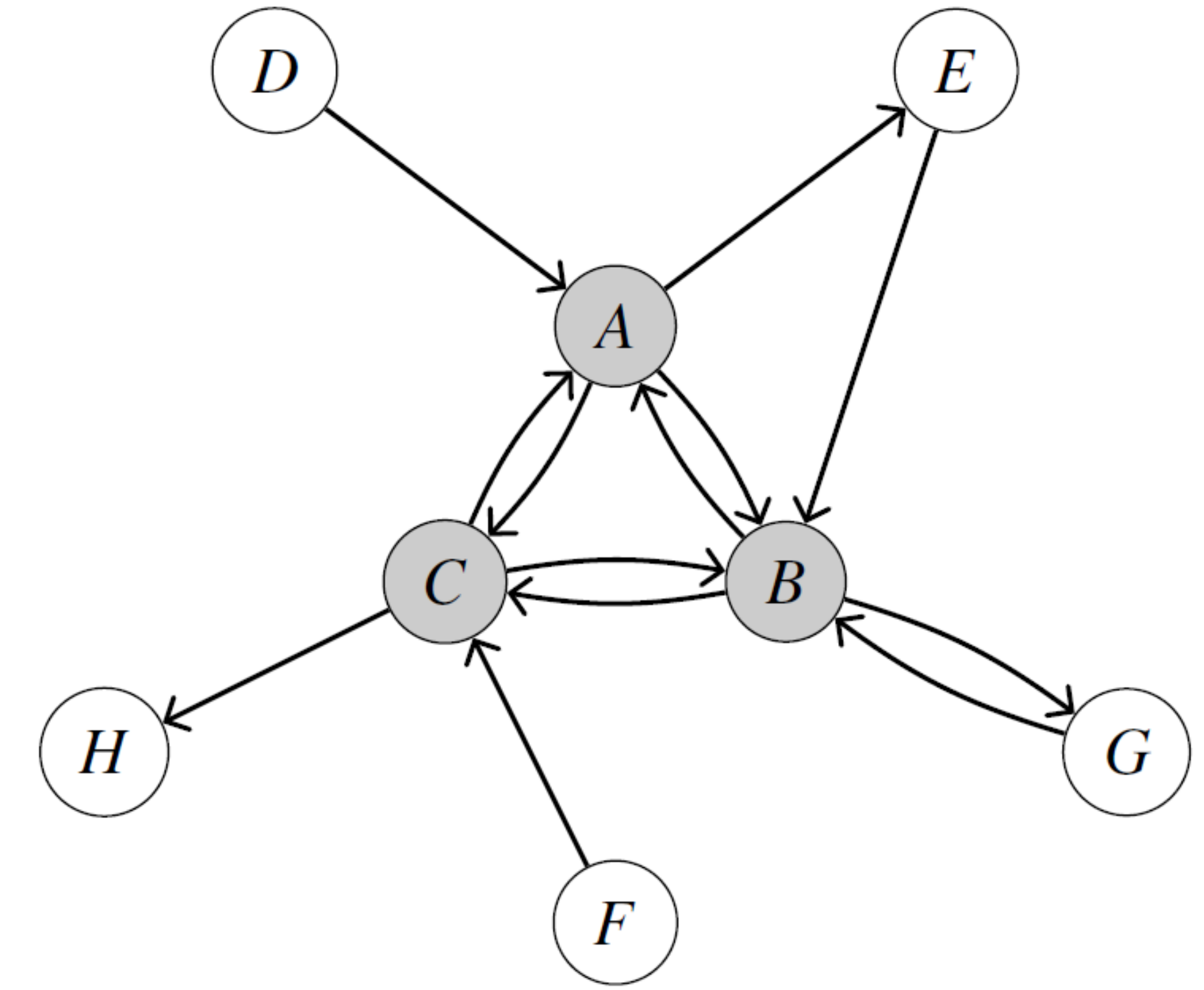}
\caption{An ideal example of core-periphery model: banks A, B, C belong to the core, and the other ones to the periphery. After ref. \cite{iman2}.}
\label{idealcp}
\end{figure}

In the literature about interbank markets (see e.g. ref. \cite{iman2}), the Core-Periphery (CP) model is a popular  axiomatic model, describing an ideal interbank network where nodes are perfectly split in two different classes: \emph{core-nodes} and \emph{periphery-nodes}.
Thus, the core and the periphery are two non-overlapping sets of vertices, defined by means of the following three axioms \cite{iman2}:
\begin{enumerate}
\item[{\bf A1}:] core banks are all bilaterally linked with each other;
\item[{\bf A2}:] periphery banks do not lend to each other;
\item[{\bf A3}:] core banks both lend to and borrow from at least one periphery bank.
\end{enumerate}
These axioms describe an ideal structure like that shown in fig. \ref{idealcp}.

Of course, real interbank networks differ from a perfect core-periphery structure as postulated by the model. Nonetheless, it is still possible to look for the optimal partition of vertices between core and periphery, i.e. the partition which minimizes the number of deviations from the CP model. In particular, three types of deviations, or `errors' can be defined, depending on which of the three axioms is violated \cite{iman2}.
For instance, if there is only one link between two banks that have been assigned to the core (instead of the postulated two links in the CP model), this is considered as one error of the first type.
And if there is one link between two periphery banks, this is considered as one error of the second type. Finally, if a core bank does not lend (borrow) to the periphery at all this is assigned as many errors as there are periphery banks.
For each of the three types of deviations, the `error score' is simply the number of errors of that type.
The total error score (for a given partition) is the sum of the scores of all three types of errors:
\begin{equation}
 \epsilon\equiv\epsilon_1+\epsilon_2+\epsilon_3\nonumber
\end{equation}
The optimal partition is the one that minimizes $\epsilon$, i.e. the one that best approximates the CP model.

We used a computational algorithm \cite{iman2} that looks for optimal solutions.
Given the resulting partition, we measured the number $N_c$ of nodes in the core and the number $N_p=N-N_c$ of nodes in the periphery. Similarly, we measured the number of links in the core as
\begin{equation}
L_c\equiv\sum_{i\neq j,\:(i,j\:\in\:core)}a_{ij}
\end{equation}
and the number of links in the periphery as
\begin{equation}
L_p=\sum_{i\neq j,\:(i,j\:\in\:periphery)}a_{ij}=L-L_c
\end{equation}
(from now on, the symbol $\sum_{i\neq j}$ indicates the two nested sums $\sum_{i=1}^N\sum_{j(\neq i)=1}^N$).
So we can distinguish between a \emph{core-connectance} and a \emph{periphery-connectance}, respectively defined as
\begin{equation}
c_{c}\equiv\frac{L_c}{N_c(N_c-1)},\quad c_{p}\equiv\frac{L_p}{N_p(N_p-1)}.
\end{equation}
The analysis of the above quantities is shown in the main text.
In section \ref{somsec:CPevol} we employ the CP model more intensively.

\section{Null models\label{somsec:nullmodels}}

The approach presented here rests upon the \emph{exponential random graphs} formalism \cite{shannon,jaynes,HL,WF,newman_expo,mylikelihood,mymethod}: given an appropriately chosen set of graphs (in what follows the \emph{grandcanonical ensemble} $\mathcal{G}$, i.e. the collection of graphs with the same number of nodes of the observed network and a varying number of links, from zero to $N(N-1)$) a probability coefficient like the following

\begin{equation}
P(A|\vec{\theta})=\frac{e^{-H(A,\:\vec{\theta})}}{Z(\vec{\theta})}
\label{pexp}
\end{equation}

\noindent is associated with each of them. In eq. (\ref{pexp}), $H(A,\:\vec{\theta})\equiv\sum_{a}\theta_a\pi_a(A)$ is the \emph{Hamiltonian} of the graph, i.e. the linear combination of topological contraints (dependent on the particular adjacency matrix, $A$) we choose to impose on the aforementioned ensemble \cite{mymethod} and the normalization constant, $Z(\vec{\theta})\equiv\sum_{A\in\mathcal{G}}e^{-H(A,\:\vec{\theta})}$, is the \emph{partition function}.

The unknown parameters can be estimated by maximizing the \emph{log-likelihood} function of the network, $\ln\mathcal{L}(\vec{\theta})=\ln P(A|\vec{\theta})$, with respect to the constraints \cite{mylikelihood,mymethod}

\begin{equation}
\frac{\partial\ln\mathcal{L}(\vec{\theta})}{\partial\theta_a}\bigg|_{\vec{\theta}^*}\equiv0,\:\forall\:a,
\end{equation}

\noindent or, equivalently, by solving

\begin{equation}
\pi_{a}(A)=\langle\pi_a\rangle(\vec{\theta}^*)\equiv\langle\pi_a\rangle^*,\:\forall\:a
\label{exp}
\end{equation}

\noindent i.e. a list of equations imposing the values of the expected constraints to be equal to the observed ones (the term ``expected'', here, refers to the weighted average taken on $\mathcal{G}$, the weights being the probability coefficients above) \cite{mylikelihood,mymethod}.

Once the numerical values $\vec{\theta}^*$ of the parameters have been determined, the expected value of any other topological quantity of interest, $X(A)$, is simply given by:

\begin{equation}
\langle X\rangle^*=\sum_{A\in\mathcal{G}}X(A)P(A|\vec{\theta}^*).
\end{equation}

However, since the expected values of the most common quantities in complex networks theory are difficult to calculate exactly, it is often necessary to rely on the linear approximation method: $\langle X\rangle^*\simeq X(\langle A\rangle^*)$, with $\langle A\rangle^*$ indicating the expected adjacency matrix, whose elements will be indicated as $\langle a_{ij}\rangle^*\equiv p_{ij}^*$. By means of the same approximation, it is also possible to calculate the standard-deviation of the quantities of interest, as described in \cite{mymethod}.

If the chosen contraints are linear in the adjacency matrix elements (i.e. of the form $\pi(A)=\sum_{i=1}^N\sum_{j(\neq i)=1}^Na_{ij}\theta_{ij}$) the expected entries become Bernoullian functions of the unknown parameters:

\begin{equation}
\langle a_{ij}\rangle= p_{ij}=\frac{e^{-\theta_{ij}}}{1+e^{-\theta_{ij}}}\equiv\frac{x_{ij}}{1+x_{ij}}.
\label{bern}
\end{equation}

The next two subsections will be devoted to the explanation of the null models considered for the present analysis.

\subsection{Directed Random Graph Model}

The \emph{Directed Random Graph} (DRG in what follows) is the most well-known null model in complex networks theory. Its Hamiltonian is composed by only one addendum, the total number of links:

\begin{equation}
H(A,\:\vec{\theta})=\alpha\:L=\sum_{i=1}^N\sum_{j(\neq i)=1}^N\alpha\:a_{ij}.
\end{equation}

Being a linear constraint, the probability coefficients have the functional form shown in eq. (\ref{bern})

\begin{equation}
p_{ij}\equiv p\equiv\frac{e^{-\alpha}}{1+e^{-\alpha}}\equiv\frac{x}{1+x}
\end{equation}

\noindent whose unknown parameter can be estimated by solving the likelihood prescription

\begin{equation}
L(A)=\langle L\rangle^*=\sum_{i=1}^N\sum_{j(\neq i)=1}^Np_{ij}^*=\sum_{i=1}^N\sum_{j(\neq i)=1}^N\frac{x^*}{1+x^*}.
\label{rgl}
\end{equation}

Eq. (\ref{rgl}) can be immediately solved to give $p^*=\frac{L(A)}{N(N-1)}=c(A)$. So, the observed connectance is nothing more that the DRG probability that two any nodes be connected. The major drawback of this null model is that only one probability coefficient, $p$, accounts for the probability connections of every pair of nodes, thus ignoring their heterogeneity. A more refined model is presented in the next subsection.

\subsection{Directed Configuration Model}

The \emph{Directed Configuration Model} (DCM in what follows) is characterized by the following Hamiltonian:

\begin{equation}
H(A,\:\vec{\theta})=\sum_{i=1}^N\left(\alpha_ik_{i}^{out}+\beta_ik_{i}^{in}\right)=\sum_{i=1}^N\sum_{j(\neq i)=1}^N\left(\alpha_i+\beta_j\right)a_{ij}
\end{equation}

\noindent (where $k_i^{in}(A)=\sum_{j(\neq i)=1}^Na_{ji}$ is the in-degree of node $i$, $k_i^{out}(A)=\sum_{j(\neq i)=1}^Na_{ij}$ is the out-degree of node $i$): it is again a linear function of the adjacency matrix elements, leading to probability coefficients of the form

\begin{equation}
p_{ij}\equiv\frac{e^{-\alpha_{i}-\beta_j}}{1+e^{-\alpha_{i}-\beta_j}}\equiv\frac{x_{i}y_j}{1+x_{i}y_j}.
\end{equation}

The likelihood prescription for the DCM becomes \cite{mymethod}

\begin{equation}
\left\{ \begin{array}{ll}
k_{i}^{in}(A)&=\langle k_i^{in}\rangle^*=\sum_{j(\neq i)}p_{ji}^* = \sum_{j(\neq i)}\frac{x_{j}^*y_{i}^*}{1+x_{j}^*y_{i}^*},\:\forall\:i \\
k_{i}^{out}(A)&=\langle k_i^{out}\rangle^*=\sum_{j(\neq i)}p_{ij}^* = \sum_{j(\neq i)}\frac{x_{i}^*y_{j}^*}{1+x_{i}^*y_{j}^*},\:\forall\:i
\end{array} \right.
\label{liksis}
\end{equation}

\noindent where the indices run from 1 to $N$. The in-degree and out-degree of a node are nothing more than the number of banks a vertex receives loans from and the number of banks a vertex lends to. Apart from considering trivial cases, the previous system can be solved only numerically.

\subsection{Reciprocal Configuration Model}
The third null model we considered is the \emph{Reciprocal Configuration Model} (RCM in what follows), where each vertex has the same number of reciprocated, out-going non-reciprocated, and in-coming non-reciprocated links as in the observed network.
In other words, the RCM incorporates not only the information about the number of (in- and outward) neighbors of a bank, but also the local reciprocity structure of each node, by means of three degree sequences defined by using the dyadic variables in eqs. (\ref{uno}-\ref{tre}) \cite{mymethod,myrec1,myrec2}:

\begin{equation}
\left\{ \begin{array}{ll}
k_{i}^{\rightarrow}(A) &= \sum_{j(\neq i)}a_{ij}^{\rightarrow}\equiv \sum_{j(\neq i)}a_{ij}(1-a_{ji}),\:\forall\:i\\
k_{i}^{\leftarrow}(A) &= \sum_{j(\neq i)}a_{ij}^{\leftarrow}\equiv \sum_{j(\neq i)}a_{ji}(1-a_{ij}),\:\forall\:i\\
k_{i}^{\leftrightarrow}(A) &= \sum_{j(\neq i)}a_{ij}^{\leftrightarrow}\equiv \sum_{j(\neq i)}a_{ij}a_{ji},\:\forall\:i
\end{array} \right.
\label{rcmvar}
\end{equation}

\noindent so that the resulting Hamiltonian becomes

\begin{equation}
H(A,\:\vec{\theta})=\sum_{i=1}^N\left(\alpha_ik_{i}^{\rightarrow}+\beta_ik_{i}^{\leftarrow}+\gamma_ik_i^{\leftrightarrow}\right).
\end{equation}

The first degree counts the number of links coming to node $i$ and not having a reciprocal partner, the second degree counts the number of links going out from node $i$ and not having a reciprocal partner and the third degree counts the number of links outgoing from, or ingoing to, node $i$ and having a reciprocal partner. The RCM imposes the above sequences as contraints on the grandcanonical ensemble. As a consequence, the likelihood condition prescribes to solve the following system \cite{mymethod}:

\begin{equation}
\left\{ \begin{array}{ll}
k_{i}^{\rightarrow}(A) &= \langle k_i^{\rightarrow}\rangle^*=\sum_{j(\neq i)}\frac{x_{i}^*y_{j}^*}{1+x_{i}^*y_{j}^*+x_{j}^*y_{i}^*+z_{i}^*z_{j}^*},\:\forall\:i\\
k_{i}^{\leftarrow}(A) &= \langle k_i^{\leftarrow}\rangle^*=\sum_{j(\neq i)}\frac{x_{j}^*y_{i}^*}{1+x_{i}^*y_{j}^*+x_{j}^*y_{i}^*+z_{i}^*z_{j}^*},\:\forall\:i\\
k_{i}^{\leftrightarrow}(A) &= \langle k_i^{\leftrightarrow}\rangle^*= \sum_{j(\neq i)}\frac{z_{i}^*z_{j}^*}{1+x_{i}^*y_{j}^*+x_{j}^*y_{i}^*+z_{i}^*z_{j}^*},\:\forall\:i
\end{array} \right.
\label{rcmeqs}
\end{equation}

Once the unknown parameters have been numerically determined, the probability coefficients $\langle a_{ij}^{\rightarrow}\rangle^*\equiv(p_{ij}^{\rightarrow})^*$, $\langle a_{ij}^{\leftarrow}\rangle^*\equiv(p_{ij}^{\leftarrow})^*$ and $\langle a_{ij}^{\leftrightarrow}\rangle^*\equiv(p_{ij}^{\leftrightarrow})^*$ can be used to calculate the expected value of all the topological properties of interest. Note that the usual degree sequences are preserved under the RCM, because

\begin{equation}
k_i^{out}=k_i^{\rightarrow}+k_i^{\leftrightarrow},\:k_i^{in}=k_i^{\leftarrow}+k_i^{\leftrightarrow}.
\end{equation}

\section{Reciprocity and dyads}
\begin{figure}[t!]
\centering
\includegraphics[width=.59\textwidth]{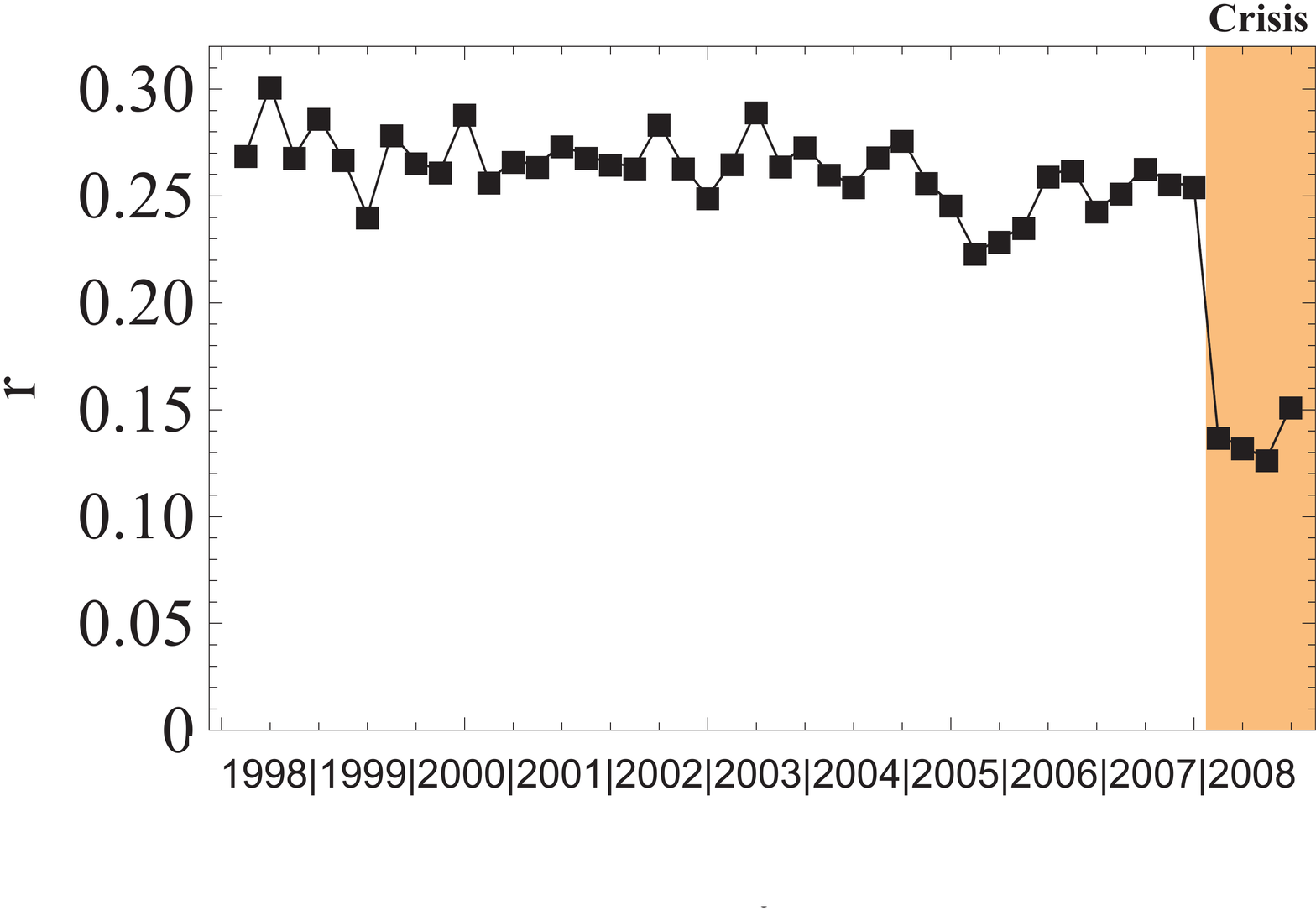}
\caption{Evolution of the reciprocity $r$ of the interbank network.}
\label{recip}
\end{figure}

The \emph{reciprocity} is the fraction of links having a reciprocal partner (i.e. a link pointing in the opposite direction) \cite{newrec}, and is defined as
\begin{equation}
r\equiv\frac{\sum_{i=1}^N\sum_{j(\neq i)=1}^Na_{ij}a_{ji}}{L}\equiv\frac{L^{\leftrightarrow}}{L}
\end{equation}
Unlike the number of vertices, the number of links, and the connectance, the reciprocity offers a clear signature of the crisis, as is evident in fig. \ref{recip}. For most of the time period it shows an essentially constant trend, with small fluctuations around an average value of approximately 0.26, but the last four periods are characterized by an impressive decrease of the reciprocity value (approximately 40\%):
they lie almost 3 sigmas away from the sample average, clearly indicating that the DIN shows an anomalously low reciprocity value in those time periods already affected by the crisis.

What about the periods before it? Just looking at fig. 2, there is no strong evidence of the upcoming event, an thus the reciprocity by itself isn't sufficiently informative about the near future. However, the statistical significance of this conclusion can only be stated after a comparison with a well defined reference, i.e. the \emph{null models} introduced in the previous section.

\subsection{The $\rho$ index}

Fig. \ref{recip} shows the effectiveness of reciprocity $r$ in characterizing the first year of the crisis. How statistically significant is the observed trend? To answer this question, let us implement the DRG and the DCM to compare the observed $r$ with its expected value:

\begin{equation}
\langle r\rangle\equiv\frac{\langle L^{\leftrightarrow}\rangle}{\langle L\rangle}=\frac{\sum_{i\neq j}p_{ij}p_{ji}}{\sum_{i\neq j}p_{ij}}
\end{equation}

\begin{figure}[t!]
\centering
\includegraphics[width=.44\textwidth,angle=270]{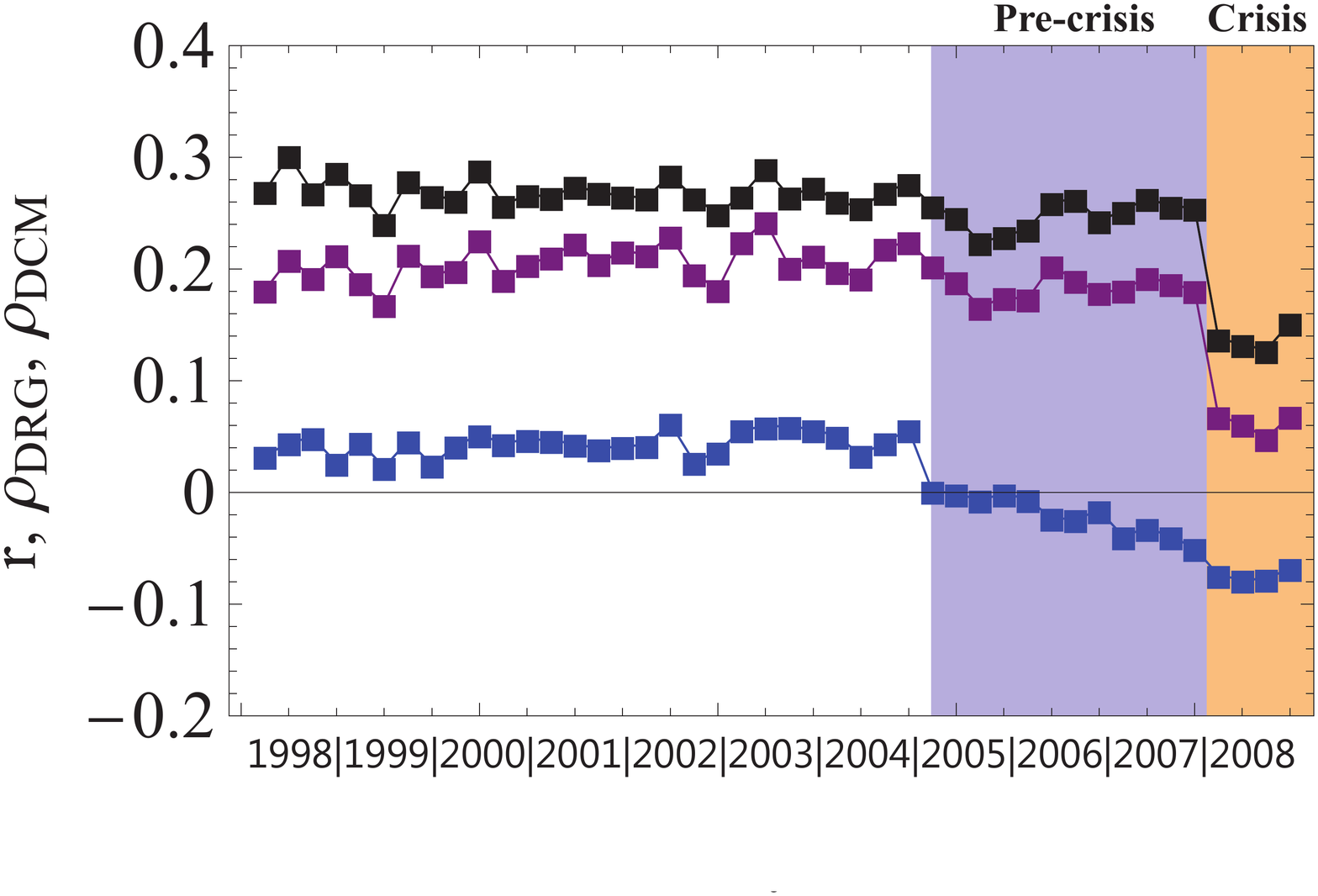}
\caption{Temporal evolution of the observed reciprocity $r$ (black) and of $\rho$ under the DRG (purple) and the DCM (blue).}
\label{errc}
\end{figure}

In order to do this, let us calculate the $\rho$ index \cite{myrec1}, defined as

\begin{equation}
\rho\equiv\frac{r-\langle r\rangle}{1-\langle r\rangle}
\end{equation}

\noindent which automatically discounts for the effects of the imposed constraints. By definition, $\rho$ ranges between 1 and $-1$: in fact, the denominator is always positive and, in magnitude, smaller than the numerator. It simply normalizes the index, not contributing to the sign of the quantity itself which, in turn, is decided only by the relative magnitude between the observed value $r$ and its expectation. A positive sign indicates a stronger than expected tendency to reciprocate whereas a negative sign, a tendency weaker than expected to establish reciprocal links. The trends of $\rho$ calculated under the DRG and the DCM are shown in fig. \ref{errc}.

The positive sign of the trend of $\rho$ under the DRG (i.e. $\rho_{DRG}$) indicates that the tendency of the network to reciprocate is stronger than expected under this model. This is intuitive by considering that $\langle r\rangle_{DRG}=p$ and that $p$ is an average of all the single pair-specific probabilities, whose numerical value coincides with the observed connectance. In fact,

\begin{eqnarray}
\langle r\rangle_{DRG}&=&\frac{\sum_{i\neq j}p^2}{\sum_{i\neq j}p}=p=\frac{\sum_{i\neq j}p_{ij}}{N(N-1)}=\frac{\langle L\rangle_{DRG}}{N(N-1)}=\nonumber\\
&=&\langle c\rangle_{DRG}
\end{eqnarray}

\noindent and, by using the likelihood prescription, $\langle c\rangle_{DRG}=c(A)$ for any matrix $A$ of the time period considered.
Given the low value of the connectance, the DRG predicts a low value for $r$ as well, such that $c<r$ for all the time-periods, thus underestimating the tendency of the links to establish mutual connections. Moreover, even if the DRG correctly describes the first year of the crisis (as evident by noting the final jump of $\rho_{DRG}$), this is exclusively due to the particular functional form of $\rho_{DRG}$ itself, being a simple, small translation (and rescaling) of $r$ towards lower values: $\rho_{DRG}=\frac{r-c}{1-c}\simeq r-c$. Thus, the discovery of patterns anticipating the crisis is trivially demanded to $r$ which is, by itself, blind to this, as already pointed out. As a result, the network seems to suddenly depart from the initial (quite stable) configuration with many reciprocated links to the crisis configuration, where the number of reciprocal links sharply diminishes.

Far more interesting is the trend of $\rho_{DCM}$. As a general comment, the network is more consistent with the DCM null rather than with the DRG one, as the smaller values of the respective $\rho$ indices show. In more detail, the DCM highlights two opposite patterns. During the first twenty-eight periods, the tendency of the network is to be reciprocated more than expected (similar to the DRG, but with the difference that $\rho_{DCM}$ presents an almost constant trend, by showing smaller fluctuations than $\rho_{DRG}$): this implies that even the specification of the entire in- and out-degree sequences is not enough to fully account for the observed reciprocity, as the positive value of $\rho_{DCM}$ witnesses. The same is valid also in the second sixteen periods, with the difference that the network inverts the tendency and tends to be less reciprocated than expected, showing an almost perfect monotonic decrease (with small, constant jumps in the value of $\rho_{DCM}$, of approximately four periods each). This clear anti-reciprocal behavior, not detected by the DRG but revealed by the DCM (i.e. not encoded in the total number of links, but partially encoded in the degree sequences) may be an early signature of the upcoming crisis, as the nodes start avoiding mutual exchanges two years before the 2008.
However, the absence of significance bounds for $\rho$ prevents us from drawing definitive conclusions about this point. Moreover, $\rho$ is a generic index, only describing the global tendency of links to reciprocate or not. This leads to the question how the single pairs of nodes behave, as discussed in the next section.

\subsection{Dyadic motifs}
\begin{figure}[t]
\centering
\includegraphics[width=.5\textwidth]{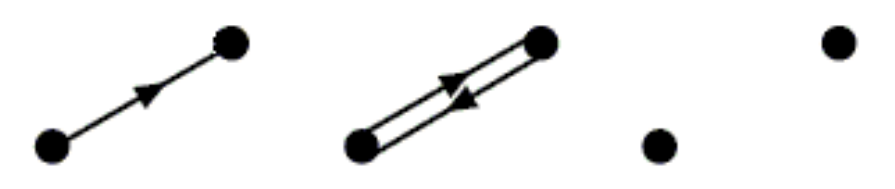}
\caption{The 3 possible binary, directed dyads.}
\label{dydy}
\end{figure}

In order to address the above points, we carried out a more detailed analysis of the reciprocity structure of the DIN, by looking at the possible \emph{dyadic motifs}, i.e. the three ways any two given nodes can be (dis)connected in a binary, directed network (see fig. \ref{dydy}).
Dyads provide detailed information about the local reciprocity structure of a network, by measuring how many single, or mutual, connections a given node has.
The statistical significance of the $\rho$ index temporal evolution can be confirmed by checking the statistical significance of the single dyads' behavior, which can in turn be measured by means of $z$-scores \cite{mymethod,myrec2}.
The number $L^{\rightarrow}$ of non-reciprocated (single) dyads, (twice) the number $L^{\leftrightarrow}$ of reciprocated (full) dyads, and (twice) the number $L^{\nleftrightarrow}$ of empty dyads are defined below, along with the corresponding $z$-scores:
\begin{equation}
L^{\rightarrow}\equiv\sum_{i\neq j}a_{ij}(1-a_{ji});\:z_{L^{\rightarrow}}=\frac{L^{\rightarrow}-\langle L^{\rightarrow}\rangle}{\sigma[L^{\rightarrow}]},
\label{uno}
\end{equation}
\begin{equation}
L^{\leftrightarrow}\equiv\sum_{i\neq j}a_{ij}a_{ji};\:z_{L^{\leftrightarrow}}=\frac{L^{\leftrightarrow}-\langle L^{\leftrightarrow}\rangle}{\sigma[L^{\leftrightarrow}]},
\label{due}
\end{equation}
\begin{equation}
L^{\nleftrightarrow}\equiv\sum_{i\neq j}(1-a_{ij})(1-a_{ji});\:z_{L^{\nleftrightarrow}}=\frac{L^{\nleftrightarrow}-\langle L^{\nleftrightarrow}\rangle}{\sigma[L^{\nleftrightarrow}]}.
\label{tre}
\end{equation}
The expected value and sigma of the dyads can be calculated analytically under both the DRG and the DCM considering that, for networks with local constraints, the dyads in a network are independent random variables \cite{mymethod}.
The temporal evolution of the dyadic $z$-scores under the DRG and the DCM is portrayed in the Main Text (note that, under the RCM, each of the three dyadic abundances is reproduced by construction, which trivially implies that all dyadic $z$-scores are zero).

The dyadic $z$-scores show the same behaviour as the $\rho$ index. The explanation lies in the analytical form of such quantites: the numerators of the $\rho$ index and the dyads are (except for a minus sign) the same. In fact

\begin{equation}
\rho=\frac{r-\langle r\rangle}{1-\langle r\rangle}=\frac{\frac{L^{\leftrightarrow}}{L}-\frac{\langle L^{\leftrightarrow}\rangle}{L}}{1-\frac{\langle L^{\leftrightarrow}\rangle}{L}}=\frac{L^{\leftrightarrow}-\langle L^{\leftrightarrow}\rangle}{L-\langle L^{\leftrightarrow}\rangle}
\end{equation}

\noindent and, considering that $L=L^{\leftrightarrow}+L^{\rightarrow}$,

\begin{equation}
L^{\rightarrow}-\langle L^{\rightarrow}\rangle=-(L^{\nleftrightarrow}-\langle L^{\nleftrightarrow}\rangle)=-(L^{\leftrightarrow}-\langle L^{\leftrightarrow}\rangle).
\end{equation}

\noindent Note that, under the RCM, all local and global dyadic properties of the real network are preserved. As a consequence, $z_{L^{\leftrightarrow}}$, $z_{L^{\rightarrow}}$ and $z_{L^{\nleftrightarrow}}$ are, by construction, zero.
For the same reason, we also have $\rho_{RCM}=0$, and the observed reciprocity structure is completely reproduced by the model.
So, the RCM fixes the dyadic motifs and can reveal patterns of self-organization \emph{between} dyads, pointing out how triads (or more numerous sets) of nodes interact.

\section{Triads}
\begin{figure}[t!]
\centering
\includegraphics[width=.99\textwidth]{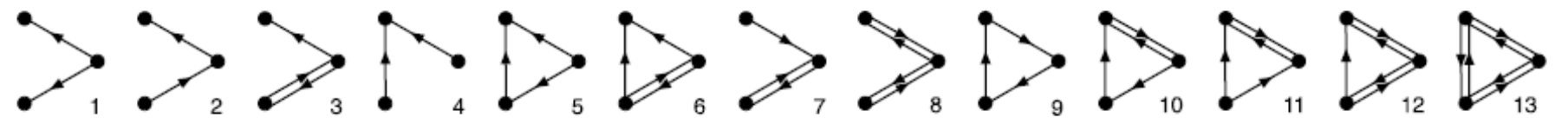}
\caption{The 13 possible triadic motifs involving three connected vertices.}
\label{mot_img}
\end{figure}

The 13 possible motifs involving three connected vertices are shown in fig. \ref{mot_img}.
The analysis of these triadic motifs has been carried out by implementing the DCM and the RCM, as shown in the Main Text. Here we provide details of that analysis, and more results relating to the DRG.

\subsection{Triadic $z$-scores}
For each motif $m=1,\dots 13$, the abundance $N_m$ (up to a constant factor $\alpha_m$ that depends on the symmetry of the particular motif, which will drop out of all measured quantities) is obtained as shown in table \ref{motable}.

\begin{table}[ht!]
\centering
\begin{tabular}{cc}
\hline\noalign{\smallskip}
$\mbox{Triadic motif}\:\:(m)$ & $\mbox{Abundance}\:\:(N_{m})$ \\
\noalign{\smallskip}
\hline
\noalign{\smallskip}
$1$ & $\sum_{i\neq j\neq k}(1-a_{ij})a_{ji}a_{jk}(1-a_{kj})(1-a_{ik})(1-a_{ki})$\\
\hline
$2$ & $\sum_{i\neq j\neq k}a_{ij}(1-a_{ji})a_{jk}(1-a_{kj})(1-a_{ik})(1-a_{ki})$\\
\hline
$3$ & $\sum_{i\neq j\neq k}a_{ij}a_{ji}a_{jk}(1-a_{kj})(1-a_{ik})(1-a_{ki})$\\
\hline
$4$ & $\sum_{i\neq j\neq k}(1-a_{ij})(1-a_{ji})a_{jk}(1-a_{kj})a_{ik}(1-a_{ki})$\\
\hline
$5$ & $\sum_{i\neq j\neq k}(1-a_{ij})a_{ji}a_{jk}(1-a_{kj})a_{ik}(1-a_{ki})$\\
\hline
$6$ & $\sum_{i\neq j\neq k}a_{ij}a_{ji}a_{jk}(1-a_{kj})a_{ik}(1-a_{ki})$\\
\hline
$7$ & $\sum_{i\neq j\neq k}a_{ij}a_{ji}(1-a_{jk})a_{kj}(1-a_{ik})(1-a_{ki})$\\
\hline
$8$ & $\sum_{i\neq j\neq k}a_{ij}a_{ji}a_{jk}a_{kj}(1-a_{ik})(1-a_{ki})$\\
\hline
$9$ & $\sum_{i\neq j\neq k}(1-a_{ij})a_{ji}(1-a_{jk})a_{kj}a_{ik}(1-a_{ki})$\\
\hline
$10$ & $\sum_{i\neq j\neq k}(1-a_{ij})a_{ji}a_{jk}a_{kj}a_{ik}(1-a_{ki})$\\
\hline
$11$ & $\sum_{i\neq j\neq k}a_{ij}(1-a_{ji})a_{jk}a_{kj}a_{ik}(1-a_{ki})$\\
\hline
$12$ & $\sum_{i\neq j\neq k}a_{ij}a_{ji}a_{jk}a_{kj}a_{ik}(1-a_{ki})$\\
\hline
$13$ & $\sum_{i\neq j\neq k}a_{ij}a_{ji}a_{jk}a_{kj}a_{ik}a_{ki}$\\
\hline
\end{tabular}
\caption{Classification and abundances (up to a symmetry factor) of the 13 triadic motifs. The three nested sums run from 1 to $N$.}
\label{motable}
\end{table}

\noindent The $z$-score for the abundance of a particular triadic motif reads
\begin{equation}
z_m=\frac{\alpha_m N_m-\langle \alpha_m N_m\rangle}{\sigma[\alpha_m N_m]}=\frac{N_m-\langle N_m\rangle}{\sigma[N_m]},
\label{ztry}
\end{equation}

\subsection{Triadic structure under the DRG}
\begin{figure}[t!]
\centering
\includegraphics[width=.6\textwidth]{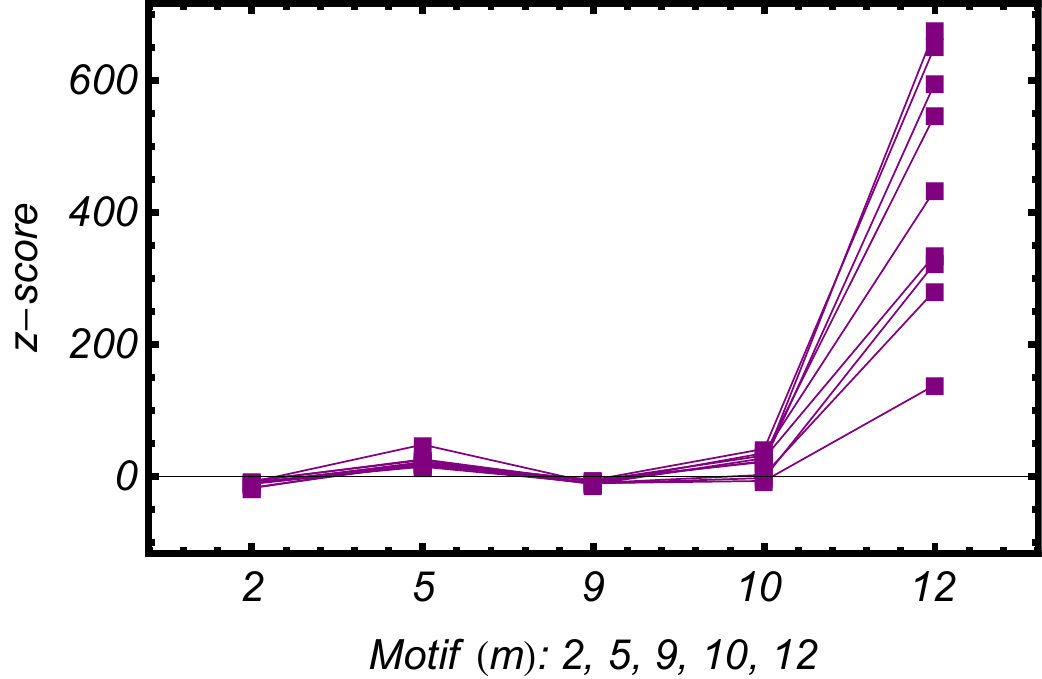}
\caption{Triadic $z$-scores for the motifs 2, 5, 9, 10, 12 in the quarterly snapshots 1, 5, 10, 15, 20, 25, 30, 35, 40, 45, under the DRG.}
\label{motDRG2}
\end{figure}

\begin{figure}[t!]
\centering
\includegraphics[width=.7\textwidth,angle=270]{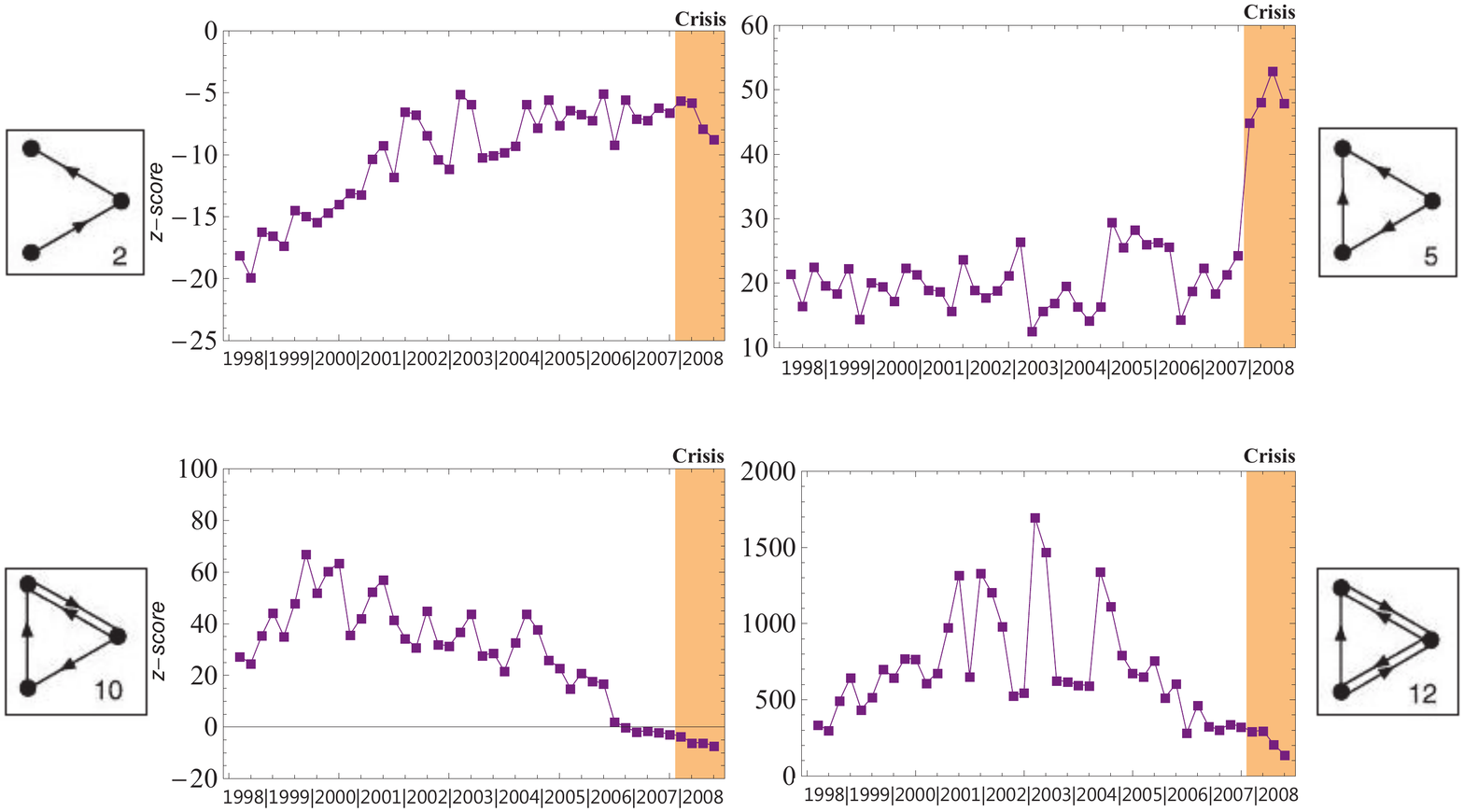}
\caption{Temporal evolution of motif 2 (top-left), motif 5 (top-right), motif 10 (bottom-left), motif 12 (bottom-right), under the DRG.}
\label{motDRG}
\end{figure}

\begin{figure}[ht!]
\centering
\includegraphics[width=.35\textwidth,angle=270]{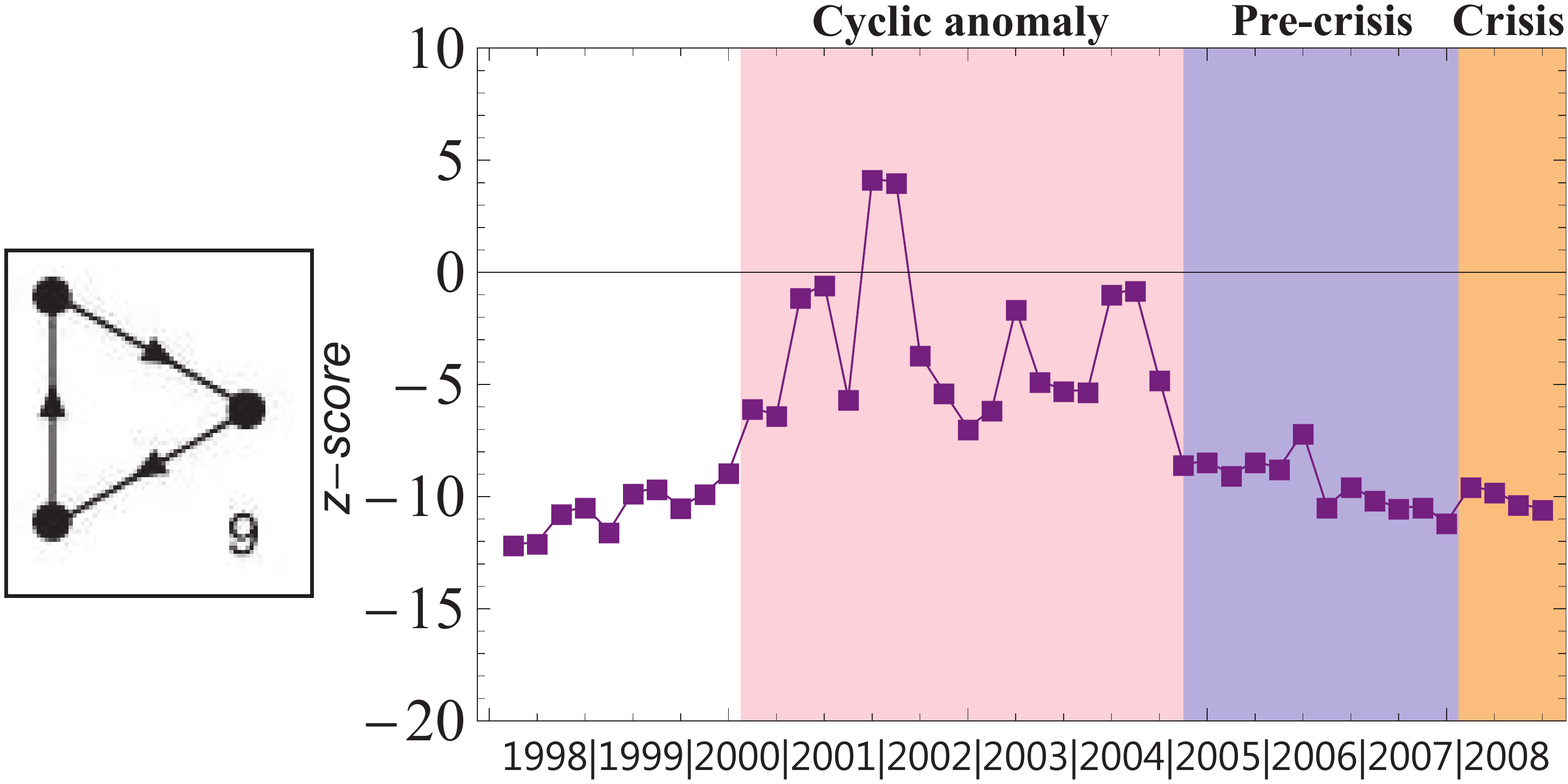}
\caption{Temporal evolution of motif 9, under the DRG.}
\label{mot_9DRG}
\end{figure}

Under the DRG, the $z$-scores are easy to calculate analytically, considering that $p_{ij}=p$, $\forall\:i\neq j$ and by using the linear approximation to calculate the standard-deviations (as shown in ref. \cite{mymethod}):

\begin{equation}
\langle N_m\rangle=T_1p^k(1-p)^{6-k},\:\sigma_{N_m}=T_2\left[kp^{k-1}(1-p)^{6-k}-(6-k)p^k(1-p)^{5-k}\right]
\end{equation}

\noindent where $k=2,\:3,\:4,\:5$ respectively for motif 2, 5 and 9, 10, 12, $T_1\equiv N(N-1)(N-2)$ is the number of distinct, directed triads and $T_2\equiv(N-2)\sqrt{N(N-1)p(1-p)}$.
The motifs considered here are the ones already considered in the Main Text: motifs 2, 5, 9, 10 and 12.
The results are shown in figs. \ref{motDRG2}, \ref{motDRG} and \ref{mot_9DRG}.

The $z$-scores under the DRG do not single out any further division in sub-periods, as shown in fig. \ref{motDRG2}: motif 12 is the only one that displays variations, but the latter correspond to a non-monotonic, oscillating temporal trend, as shown in fig. \ref{motDRG}.
Motifs 2, 10 and 12 do not even signal the anomalous period of the crisis, as evident by looking at fig. \ref{motDRG}: in fact, no significant departure from the previous periods-trend is appreciable. In this respect, the only useful trend is provided by motif 5, which shows an evident increase just before the beginning of the critical period. However, no evidence of a pre-crisis period is detectable in any of the four considered motifs, confirming the limited use of the DRG in providing useful predictions.
So the above triadic motifs calculated under the DRG can at best confirm the same unpredictability as the dyadic ones (some triadic patterns undergo a sudden change in 2008), and at worst show fluctuations which are so large during the entire period, that it is difficult to tell whether any significant change is taking place at all at the onset of the crisis.

Only motif 9 shows a more informative temporal evolution and partly confirms (even if with much less significance) the results we found for the same motif under the DCM and the RCM (see main text), as fig. \ref{mot_9DRG} shows. Even if neither the crisis, nor the pre-crisis period are detected, the anomaly in the `cyclic anomaly' period is still visible, appearing as a global increase of the $z$-scores' values before coming back to an almost constant value in the last temporal periods $t_{30}-t_{44}$.

\subsection{Triadic structure under the DCM}
\begin{figure}[t!]
\centering
\includegraphics[width=.6\textwidth]{Supplementary_Figure_5}
\includegraphics[width=.6\textwidth]{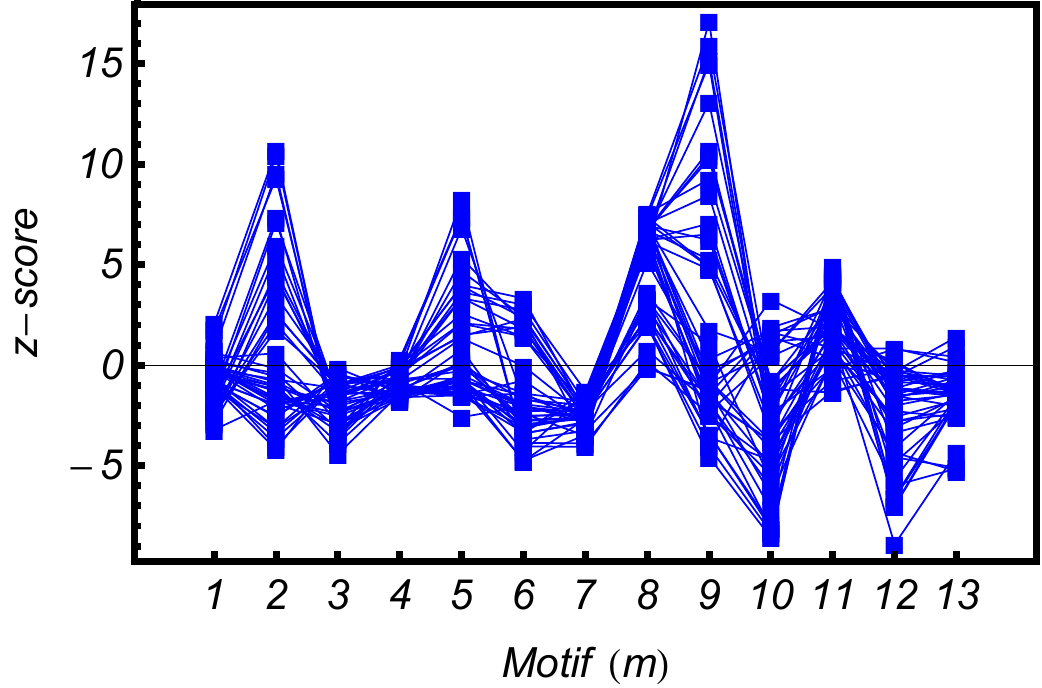}
\caption{Triadic $z$-scores for all 44 quarters, under the DCM.}
\label{mot_all}
\end{figure}

Motifs' profiles under the DCM represent one of the most significant results of the whole analysis, unambiguously showing the evolution of the DIN, as clear by looking at the forty-four motifs' profiles plotted together in fig. \ref{mot_all}. The apparent disorder resulting from plotting all the forty-four profiles together hides the four, stationary sub-profiles shown in the main text. We will discuss each of the sub-periods in turn.

In the first subperiod (covering the time-periods between $t_1$ and $t_{10}$ - see main text), the DCM seems to explain quite well the motifs' profiles (almost all the $z$-scores lie between $z=+3$ and $z=-3$): the only outlier is motif 8, whose abundance is underestimated by the DCM (the $z$-score is positive).  Looking back at fig. \ref{mot_img} this means that, even if the total number of reciprocal and empty dyads is consistent with the DCM prediction in this subperiod, the abundance of triads composed by two reciprocated dyads and one empty dyad is still underestimated by it.
This suggests that the simple topological information about the number of neighbors is still not sufficient to account for this kind of dyadic interactions and more precise information about the nodes' local reciprocity structure is needed.
In the second subperiod (between $t_{11}$ and $t_{18}$), even if motif 8 is again underestimated, the biggest, evident discrepancy is between the abundance of motif 9 (rising with respect to the first ten years) and its expected value, while the other motifs remain quite stable. Motif 9 is a complete triad of single dyads. Its underestimation could be due to the presence of genuine self-organization patterns at the triadic level rather than the reciprocity structure being ignored. As in the previous case, the DCM does not account for them.
The third subperiod (between $t_{19}$ and $t_{40}$) shows several motifs becoming more significant than in the previous subperiods: motif 2 and motif 5 become overrepresented (i.e. underestimated by the DCM) and motifs 10 and 12 become underrepresented (i.e. overestimated by the DCM). In the fourth subperiod (between $t_{41}$ and $t_{44}$), these patterns become even more pronounced. The increasing divergence of these motifs from the DCM null could be another signature of the crisis, also by direct comparison with the dyadic abundance: as the number of reciprocal dyads is overestimated in the year 2008, so motif 10 and 12 are, both defined by (at least) one reciprocal dyad; as the number of single dyads is underestimated in the year 2008 (becoming only marginally consistent with the DCM prediction) so motif 2 and 5 are, both defined by (at least) two single dyads.

\subsection{Triadic structure under the RCM}
\begin{figure}[t!]
\centering
\includegraphics[width=.6\textwidth]{Supplementary_Figure_5}
\includegraphics[width=.6\textwidth]{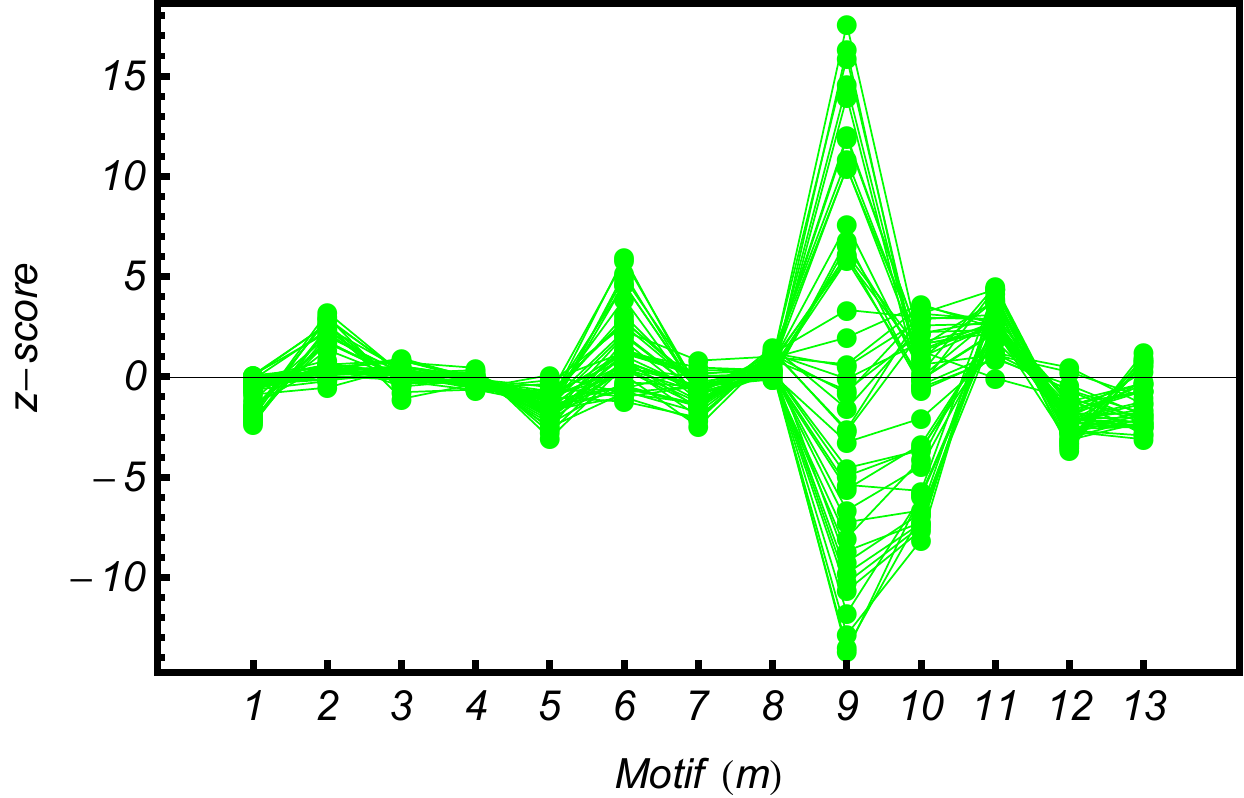}
\caption{Triadic $z$-scores for all 44 quarters, under the RCM.}
\label{mot_all2}
\end{figure}

What clearly emerges from the previous analysis is the fundamental role of the reciprocity structure of the DIN in describing both the crisis and the pre-crisis period. The DCM only partially accounts for the dyadic and the triadic structure of the network, highlighting the emergence of patterns not encoded in the degree sequences.
In order to discover the presence of higher-order patterns, not encoded in the dyadic structure, the next step is to fix this kind of topological information beforehand as well.
In other words, to disentangle dyadic effects from a genuine organization at the triadic level, we need to wash out the information about the reciprocity itself, by including it in our null model from the start.
This results in the RCM with figure \ref{mot_all2} showing the associated triadic $z$-scores.
It is clear at a glance that now, except motifs 9 and 10, all motifs are approximately consistent with the null model. This means that, after controlling for the dyadic patterns already identified, motifs 9 and 10 still emerge as strongly significant triadic building blocks, irreducible to a combination of dyads.

The temporal evolution of these triadic $z$-scores (see main text) reveals that, out of the four motifs that are significant under the DCM in the third (`pre-crisis' phase) subperiod. Only motif 10 retains the same significance under the RCM, signalling that the information about the reciprocity is not sufficient to explain the abundance of this profile. Just as for motif 9, which sharply inverts its trend, motif 10 is composed of two single dyads and one reciprocal dyad, closing a triangle loop: this seems to confirm the additional presence of non-trivial third-order correlations. In other words, the network evolves from configurations with an exceptional, unexplained abundance of unreciprocated triangular loops to configurations where nodes strongly prefer avoiding them. The fourth (`crisis') subperiod is, again, a further evolution of the third one, showing motifs 9 and 10 evolving towards more strongly significant values.
So, the first year of the crisis seems to be characterized by the strong absence of triangle interactions, while all the other patterns seem to be compatible with the RCM prediction. The evolution of motif 9 (see main text) confirms this, clearly indicating that a sort of triadic self-organization indeed exists and plays a fundamental role in shaping the interbank exchanges.

\section{Evolution of the core-periphery structure\label{somsec:CPevol}}

In this section we describe our analyses of the evolution of the core-periphery structure of the DIN.
As described in section \ref{somsec:cp}, we first looked for the the optimal partition of vertices providing the closest approximation to the CP model.
As for the dyadic and triadic structure, we are interested in understanding whether the observed core-periphery structure is statistically significant, or whether it can be explained merely in terms of the local topological properties.

\subsection{Error score}
To this end, we first studied whether the error score $\epsilon$ measured on the real network (given the optimal partition) is consistent with the value $\langle \epsilon\rangle$ (given the same partition) expected under the null models considered so far.
Given a null model, we therefore define the $z$-score
\begin{equation}
z_{\epsilon}=\frac{\epsilon-\langle\epsilon\rangle}{\sigma[\epsilon]}
\end{equation}
measuring by how many standard deviations the observed divergence from the ideal CP model differs from its expectation.
As we now show, it is possible to evaluate $z_{\epsilon}$ analytically.

Generally, the optimal partition does not contain errors of the third type, since the latter are severely punished (See section \ref{somsec:cp} for the definition).
As a consequence, it is sensible and easier to restrict our attention to the errors of the first and second type, which measure the deviation from a `relaxed' CP model without the third axiom.
If $V_c\equiv N_c(N_c-1)$ and $V_p\equiv N_p(N_p-1)$ denote the \emph{volume} (number of possible links) of the core and periphery respectively, we consider the following simplified error score:
\begin{equation}
\epsilon\equiv\epsilon_1+\epsilon_2\equiv V_c-L_c+L_p
\end{equation}
where $V_c-L_c$ is the number of missing (with respect to the CP model) links in the core and $L_p$ is the number of extra (with respect to the CP model) links in the periphery. In an ideal core-periphery model we would have $L_c=V_c$ and $L_p=0$, so that $\epsilon=0$.
Using the expression above, we can write
\begin{equation}
\langle\epsilon\rangle=V_c-\langle L_c\rangle+\langle L_p\rangle
\end{equation}
Moreover, it is easy to check that
\begin{equation}
\sigma^2[ \epsilon]=\sigma^2[L_c]+\sigma^2[L_p]
\end{equation}
Taken together, these expression yield the desired analytical formula
\begin{equation}
z_{\epsilon}=\frac{\epsilon-\langle\epsilon\rangle}{\sigma[\epsilon]}=\frac{(L_p-L_c)-\langle L_p-L_c\rangle}{\sqrt{\sigma^2[L_c]+\sigma^2[L_p]}},
\label{errfor}
\end{equation}
Equation (\ref{errfor}) shows that the fundamental variable appearing in the error definition is the difference between the links in the core and those in the periphery.

\begin{figure}[t]
\centering
\includegraphics[width=.44\textwidth,angle=270]{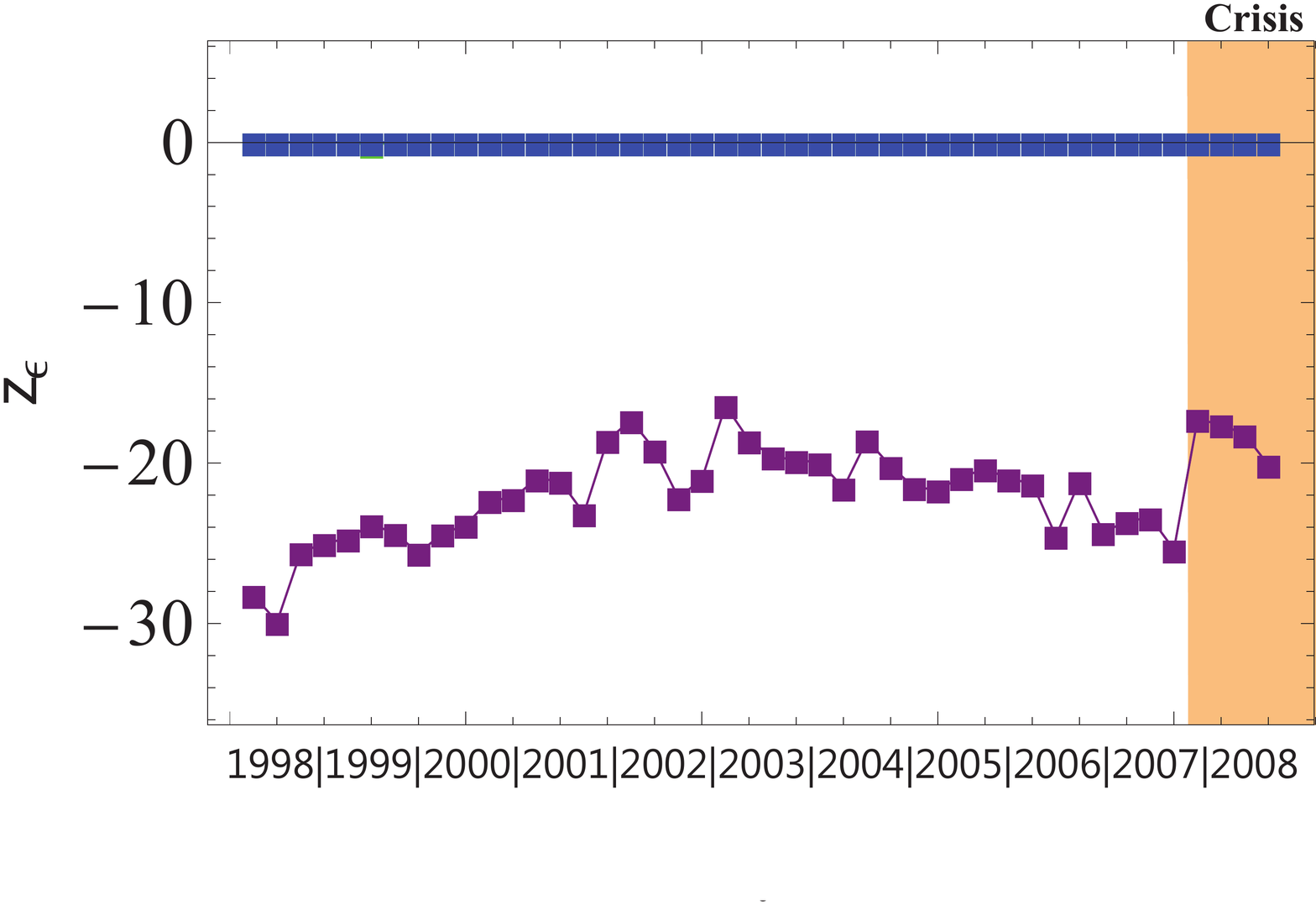}
\caption{$z$-score for the error score $\epsilon$ under the DRG (purple), the DCM (blue) and the RCM (green).}
\label{imanz}
\end{figure}

Armed with the above result, we employ the DRG, the DCM and the RCM and evaluate the corresponding $z$-scores.
The results are shown in fig. \ref{imanz}.
Using the DRG, we find that the observed core-periphery structure is strongly significant, as the DRG predicts a much larger expected error score resulting in larger negative values of $z_{\epsilon}$.
By contrast, using the DCM and RCM (which both give zero $z_{\epsilon}$), we find that the real network is as close to the ideal CP model as are null models.
The reason for $z_{\epsilon}$ being almost identically zero is that, as can be easily checked, all the possible rewiring moves that preserve the degrees of vertices \emph{exactly} also preserve the number of errors of the first and second type (note that this would imply $\langle\epsilon\rangle=0$ and $\sigma[\epsilon]$, making $z_{\epsilon}$ undefined).
Our different approach, where degrees are preserved \emph{on average} (see section \ref{somsec:nullmodels}), allows for errors not to be preserved exactly in each single realization of the network.
This implies that $\sigma[\epsilon]>0$, making $z_{\epsilon}$ properly defined.
Still, we find that $z_{\epsilon}\approx 0$, meaning that $\langle\epsilon\rangle$ is extremely close to zero.
In other words, the observed core-periphery structure, as measured from the error score, is not a genuinely higher-order property, since it is simply explained by the degrees  of vertices (for a discussion of this result see also \cite{iman2}).

\subsection{Density contrast}
To double-check the above results, we introduced an alternative measure of the strength of core-periphery structure, namely the \emph{density contrast} defined as
\begin{equation}
\Delta c\equiv c_{c}-c_{p}=\frac{L_{c}}{N_{c}(N_{c}-1)}-\frac{L_{p}}{N_{p}(N_{p}-1)}=\frac{L_{c}}{V_{c}}-\frac{L_{p}}{V_{p}}
\end{equation}
i.e., the difference between the link density in the core and the link density in the periphery (see section \ref{somsec:cp}).
This quantity is a very intuitive measure of the excess core density and, unlike the error score, is not trivially preserved by rewiring moves that preserve degrees, either exactly or on average.
The $z$-score for the density contrast is
\begin{equation}
z_{\Delta c}=\frac{\Delta c-\langle\Delta c\rangle}{\sigma[\Delta c]}=\frac{\left(\frac{L_{c}}{V_{c}}-\frac{L_{p}}{V_{p}}\right)-\left(\frac{\langle L_{c}\rangle}{V_{c}}-\frac{\langle L_{p}\rangle}{V_{p}}\right)}{\sqrt{\frac{\sigma_{L_{c}}^2}{V_{c}^2}+\frac{\sigma_{L_{p}}^2}{V_{p}^2}}}.
\end{equation}

\begin{figure}[t]
\centering
\includegraphics[width=.44\textwidth,angle=270]{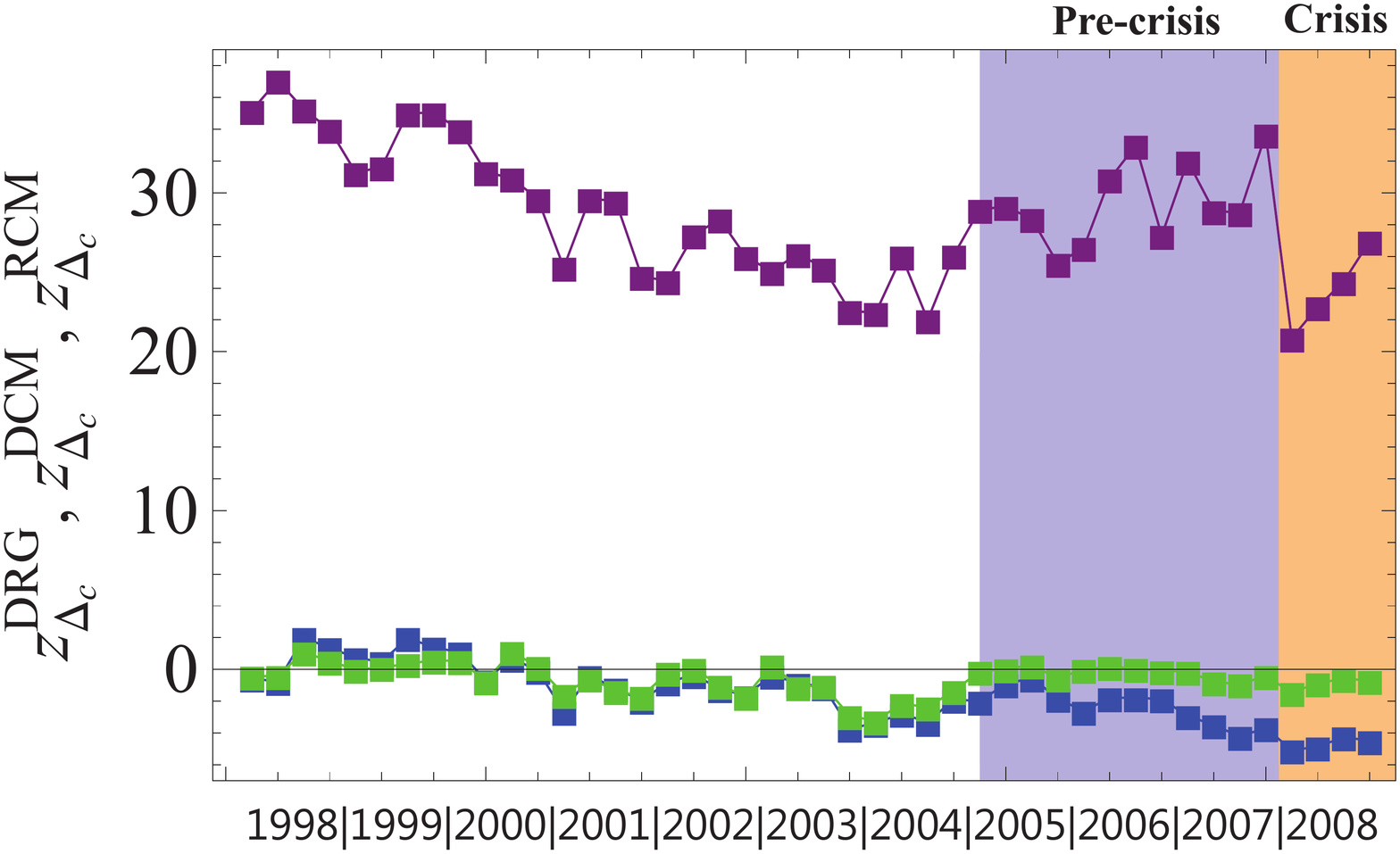}
\caption{$z$-score for the density contrast $\Delta c$ under the DRG (purple), the DCM (blue) and the RCM (green).}
\label{somerrc3}
\end{figure}

The evolution of $z_{\Delta c}$ under the three null models is shown in fig. \ref{somerrc3}.
Clearly, the DRG prediction underestimates the contrast. This is easily understandable by looking back at fig. 1 in the Main Text: the only information the DRG uses is the global connectance, thus predicting a more homogeneous structure and a smaller difference between the two zones' density than observed.
However, the first year of the crisis shows a jump towards smaller values of the $z$-score as if, exactly as pointed out by the reciprocity index $\rho$, in 2008 the network became sparser and actually more homogeneous (less difference between a ``suspected'' core and periphery), thus being in better agreement with the DRG.

The DCM and the RCM predict the same density contrast for most of the time period. The two models start deviating from each other approximately around period $t_{30}$, when the ``pre-crisis'' period starts: in particular, the DCM does not correctly account for the observed $\Delta c$, especially in the last periods (approximately two years, from $t_{37}$ to $t_{44}$), when the crisis spreads. Here, the trend of the $z$-score highlights that the DIN structure is actually more homogeneous than predicted by the DCM: the network, as it evolves towards the crisis, seems to actually lose a sort of internal structure, based on the different density of links in the core and periphery areas. Moreover, this information seems to be again encoded into the reciprocity structure of the network as the the RCM, which accounts for this, correctly explains the density contrast. This sheds new light on the evolution of the reciprocity itself, as pointed out in the conclusions.

\subsection{The distribution of $L_p-L_c$}

The $z$-scores have a well defined statistical significance levels only in the case of normally-distributed variables. We tested this assumption only for a single time period ($t_{8}$), by proceeding in the following way. We implemented the DCM numerically for the chosen time period by generating 50.000 binary, directed matrices, according to the DCM rule

\begin{displaymath}
\left\{ \begin{array}{ll}
a_{ij}=1, & \mbox{if}\:u\:\sim U[0,1]\leq p_{ij}^*=\frac{x_{i}^*y_{j}^*}{1+x_{i}^*y_{j}^*}\\
a_{ij}=0, & \mbox{else}
\end{array} \right.
\end{displaymath}

\noindent i.e. by extracting a real number uniformly distributed between 0 and 1 and comparing it with $p_{ij}^*$ (whose numerical value was computed according to the maximum of the likelihood procedure), for each entry of the adjacency matrix, $a_{ij}$. Then, for each matrix $M$ belonging to this numerically-generated grandcanonical ensemble, $\mathcal{M}$, we also calculated the distribution of the random variable $L_{p}-L_{c}$, its arithmetic mean,

\begin{equation}
\overline{L_{p}-L_{c}}=\frac{\sum_{M\in\mathcal{M}}L_{p}(M)-L_{c}(M)}{|\mathcal{M}|}
\label{f1}
\end{equation}

\noindent and its standard deviation,

\begin{equation}
\sigma^2[L_{p}-L_{c}]=\overline{(L_{p}-L_{c})^2}-\left(\overline{L_{p}-L_{c}}\right)^2.
\label{f2}
\end{equation}

\begin{figure}[t!]
\centering
\includegraphics[width=.6\textwidth]{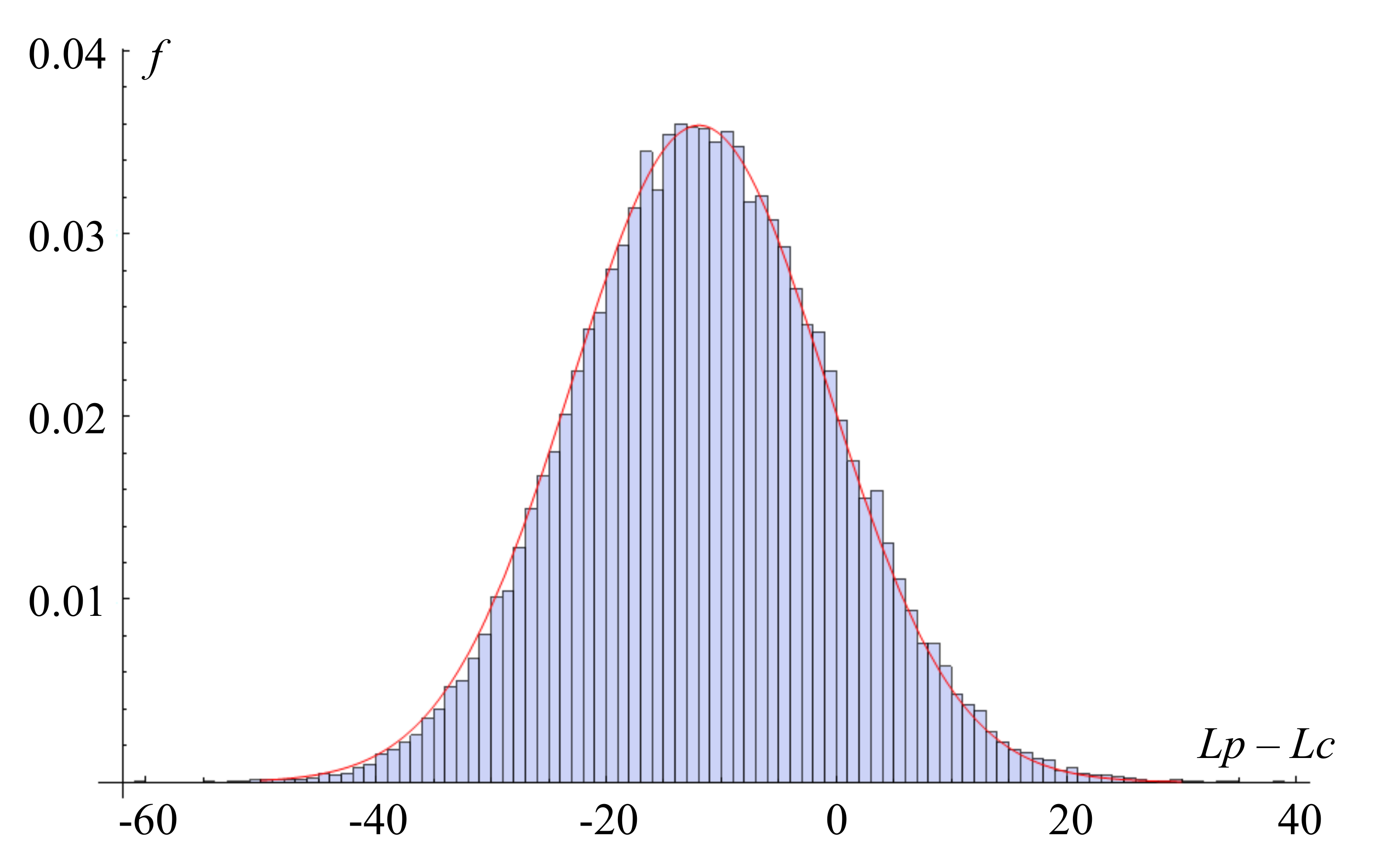}
\qquad\qquad
\includegraphics[width=.6\textwidth]{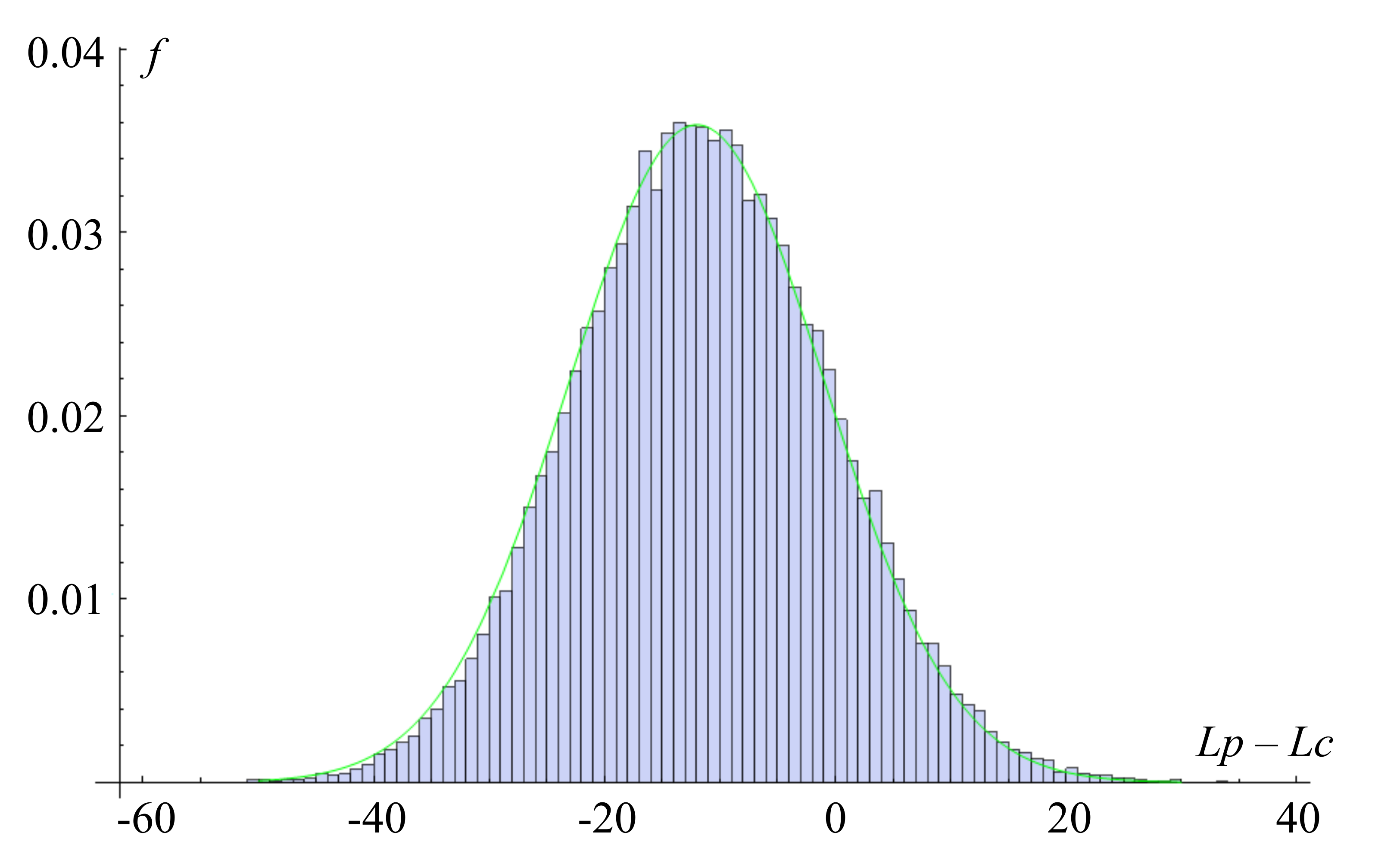}
\caption{Histogram showing the distribution of $L_p-L_c$ on $\mathcal{M}$ (a numerically-generated grandcanonical ensemble according to the DCM probability coefficients and composed by 50.000 matrices), Gaussian fit with parameters $\mu=\overline{L_{p}-L_{c}}$ and $\sigma^2=\overline{(L_{p}-L_{c})^2}-\left(\overline{L_{p}-L_{c}}\right)^2$ (top, red) and Gaussian fit with parameters $\mu=\langle L_{p}-L_{c}\rangle$ and $\sigma^2=\langle(L_{p}-L_{c})^2\rangle-\langle L_{p}-L_{c}\rangle^2$ (bottom, green).}
\label{g1}
\end{figure}

The results are shown in fig. \ref{g1}. The histograms both show the same distribution of $L_{p}-L_{c}$: we simply added two fits. The red fit is a Gaussian fit with parameters $\mu=\overline{L_{p}-L_{c}}$ and $\sigma^2=\overline{(L_{p}-L_{c})^2}-\left(\overline{L_{p}-L_{c}}\right)^2$, i.e. the same as in eqs. (\ref{f1}) and (\ref{f2}). The green fit is again a Gaussian, with parameters $\mu=\langle L_{p}-L_{c}\rangle$ and $\sigma^2=\langle(L_{p}-L_{c})^2\rangle-\langle L_{p}-L_{c}\rangle^2$, i.e. those analytically calculated by means of the maximum of the likelihood procedure (and implemented for the time-period $t_8$).
What we observe is a substantial agreement between the distribution of $L_p-L_c$ calculated on the numerically-generated grandcanonical ensemble and the two Gaussian fits, thus confirming the usual statistical interpretation of the $z$-scores.

\section{Akaike's Information Criterion (AIC)}

For the analysis of the DIN we have used three null models: the DRG, the DCM and the RCM. Which of these is the `best' null model in terms of an optimal trade-off between parsimony and ability to replicate the data?
In order to answer this question, we use the \emph{Akaike's Information Criterion} (AIC in what follows) and the \emph{Akaike weights} \cite{model_selection}. These techniques have been recently used in the analysis of other networks \cite{mydistances}.

For each snaphot $t$ and for each model, AIC is defined as the difference between (twice) the number $K$ of parameters and the log-likelihood at its maximum:

\begin{equation}
AIC^t(\vec{\theta}^*)\equiv 2K^t-2\ln\mathcal{L}^t(\vec{\theta}^*).
\end{equation}
The model with the smallest value of $AIC^t$ achieves the optimal balance between explanatory power (large log-likelihood) and parsimony (small numer of parameters) \cite{model_selection}.
Across all snaphots, we found that the DRG is never the best model, while the DCM and RCM sometimes compete.
To better discriminate between two competing models, the \emph{Akaike weights} can be used.
Given two models $1$ and $2$, we can calculate the two values (for each time period)

\begin{equation}
\Delta_i^t\equiv AIC_i^t-\min\{AIC_1^t,\:AIC_2^t\},\:i=1,\:2
\end{equation}

\noindent to define the Akaike weights \cite{model_selection} as

\begin{equation}
w_1^t\equiv\frac{e^{\frac{-\Delta_1^t}{2}}}{e^{\frac{-\Delta_1^t}{2}}+e^{\frac{-\Delta_2^t}{2}}},\:w_2^t\equiv\frac{e^{\frac{-\Delta_2^t}{2}}}{e^{\frac{-\Delta_1^t}{2}}+e^{\frac{-\Delta_2^t}{2}}}.
\end{equation}
By definition, $w_1+w_2=1$. A value $w_i\approx 1$ implies that model $i$ strongly outperforms the other model, which can therefore be discarded. Values $w_1\approx w_2\approx 1/2$ imply that both models are very similar, and neither can easily be discarded.

\begin{figure}[t!]
\centering
\includegraphics[width=.44\textwidth,angle=270]{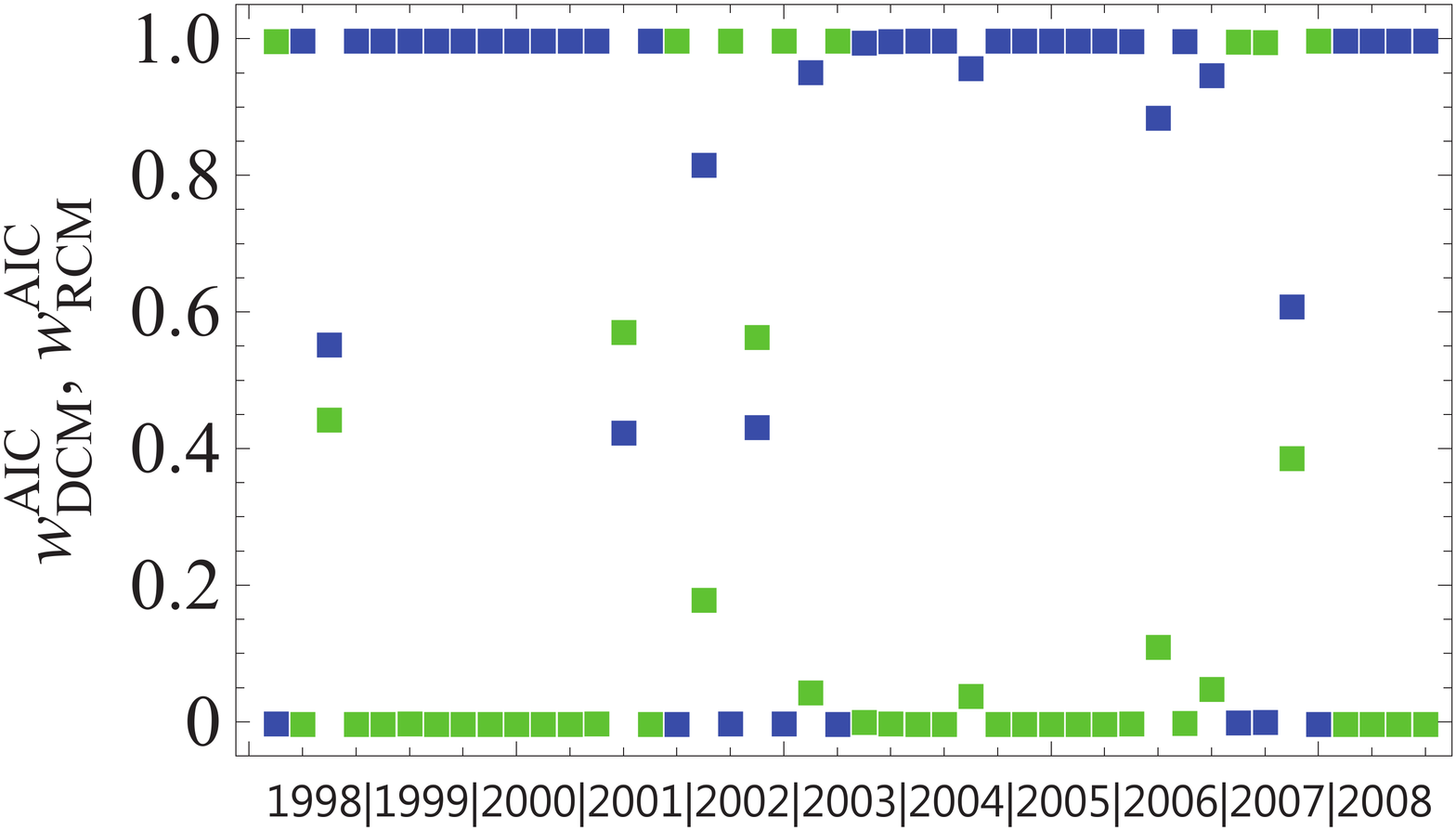}
\caption{Temporal evolution of Akaike weights for the DCM (blue) and the RCM (green).}
\label{aicw}
\end{figure}

Our results are shown in fig. \ref{aicw} for all shapshots.
For four snapshots the two models compete (having the weights near the central value of 0.5), so that they should both be retained and some more refined form of multimodel inference would be needed (the so-called \emph{multimodel average} \cite{model_selection}). However, for most of the remaining snapshots the DCM outperforms the RCM, which is therefore less effective in explaining the observed topology. Given the extreme values of the Akaike weights, AIC seems to classify RCM as an overfitting model most of the time, pointing out that the correct amount of information to use is encoded in the degree sequences only and that all the remaining higher-order structures (dyadic and triadic structure) should be considered as non trivial patterns revealing the self-organization of the DIN.

So, even if the topological information introduced by adding the local reciprocity structure to the contraints would seem not to be excessive, most of the time AIC classifies it as redundant and suggests that the optimal level of description is the one achieved by controlling for the in- and out-degree sequences, confirming that all higher-order patters starting from the dyadic ones should be regarded as significant. Clearly, in order to filter out the dyadic effect from the triadic abundances, and select only the triads which cannot be explained in terms of a combination of dyads, the RCM remains the model to use.

\section{Entropy}
In order to measure how effective the chosen constraints are in `narrowing' the ensemble of networks around the observed configuration, we calculated the Shannon entropy of the probability distribution induced by each null model
 \cite{shannon,jaynes}.
To obtain comparable results across the three null models, we normalized all entropies between 0 and 1, considering that the DRG and the DCM predict probability coefficients for the directed pairs while the RCM predicts probability coefficients for the dyads.
This normalization results in the following definition:

\begin{equation}
S_{NM}\equiv-\frac{\sum_{A\in\mathcal{G}}P_{NM}(A)\log_2 P_{NM}(A)}{N(N-1)},\:NM=DRG,\:DCM,\:RCM.
\end{equation}

\begin{figure}[t!]
\centering
\includegraphics[width=.44\textwidth,angle=270]{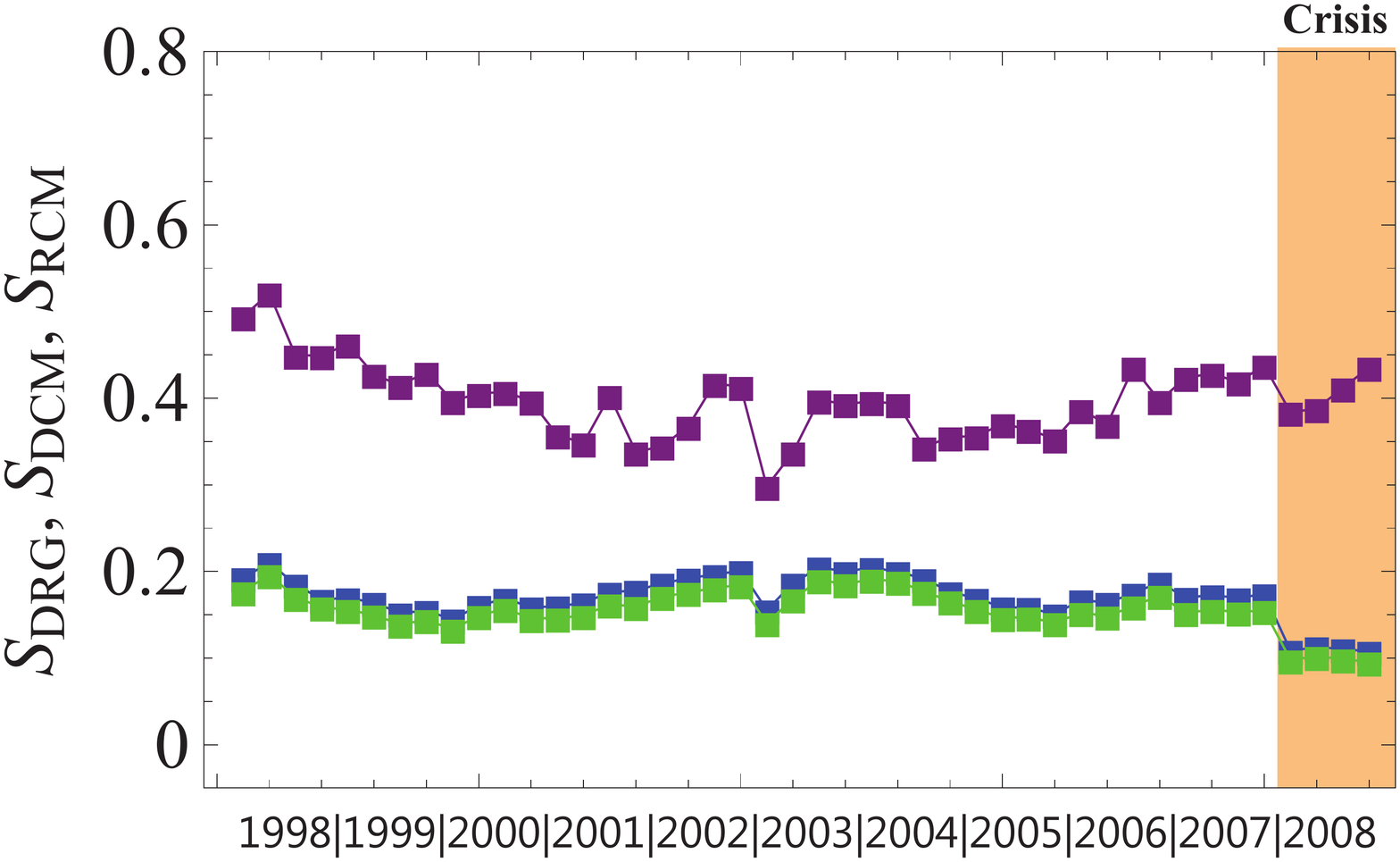}
\caption{Entropy for the three null models DRG (purple), DCM (blue) and RCM (green).}
\label{entro}
\end{figure}

The result is shown in fig. \ref{entro}. The same observations made about the connectance are also valid for the DRG entropy: the magnitude of the change between $t_{40}$ and $t_{41}$ is of the same (or even lower) order as the previous ones, so that it would be difficult to detect the crisis in terms of an anomalous behavior of $S_{DRG}$. On the other hand, both the DCM and the RCM entropies show a clear (even if not dramatically large) jump between 2007 and 2008, somehow indicating the onset of the crisis.
However, there is no clear indication, neither in the DRG prediction nor in the DCM prediction, of a pre-crisis period. This seems to indicate that the clues of the upcoming instability are detectable neither in the degree sequences, nor in the local reciprocity structure, but in higher-order statistics (triadic motifs and core-periphery division), once the dyadic structure is kept fixed (degree sequences).
Also note the small difference between $S_{DCM}$ and $S_{RCM}$. While in going from the DRG to the DCM there is clearly a large information gain, there is only a much smaller gain in going from the DCM to the RCM. This explains why AIC most of the time indicates that the information gained by the RCM over the DCM is not enough to justify the introduction of the additional model parameters.

\section*{Acknowledgements}

D. G. acknowledges support from the Dutch Econophysics Foundation (Stichting Econophysics, Leiden, the Netherlands) with funds from beneficiaries of Duyfken Trading Knowledge BV, Amsterdam, the Netherlands. This work was also supported by the Netherlands Organization for Scientific Research (NWO/OCW).


\begin{thebibliography}{99}

\bibitem{may0}
May, R. M., Levin, S. \& Sugihara, G. Ecology for bankers, \emph{Nature} {\bf 451}, 893-895 (2008).

\bibitem{may1}
Haldane, A. G. \& May, R. M. Systemic risk in banking ecosystems, \emph{Nature} {\bf 469}, 351-355 (2011).

\bibitem{subprime}
Longstaff, F. A. The subprime credit crisis and contagion in financial markets, \emph{J. Financ. Econ.} {\bf 97}, 436-450 (2010).

\bibitem{science}
Schweitzer, F., Fagiolo, G., Sornette, D., Vega-Redondo, F., Vespignani, A. \& White, D. R. Economic networks: the new challenges, \emph{Science} {\bf 325}, 422-425 (2009).

\bibitem{newman2010}
Newman, M. E. J. \emph{Networks: An Introduction} (Oxford University Press, Oxford, 2010).

\bibitem{goyal2009}
Goyal, S. \emph{Connections: An Introduction to the Economics of Networks} (Princeton University Press, Princeton, 2009).

\bibitem{jackson}
Jackson, M. O. \emph{Social and Economic Networks} (Princeton University Press, Princeton, 2010).

\bibitem{iman}
van Lelyveld, I. \& Liedorp, F. R. Interbank contagion in the Dutch banking sector: a sensitivity analysis, \emph{Int. J. Central Banking} {\bf 2}, 99-132 (2006).

\bibitem{iman2}
van Lelyveld, I. \& in't Veld, D. Finding the core: network structure in interbank markets (2012) Dutch National Bank Working Paper. Available at:\\ $http://www.dnb.nl/en/publications/dnb-publications/dnb-working-papers-series/dnb-working-papers/working-papers-2012/dnb276178.jsp$ (Accessed: 06/11/2013).

\bibitem{iman3}
Liedorp, F. R., Medema, L., Koetter, M., Koning, R. H. \& van Lelyveld, I. Peer monitoring or contagion? Interbank market exposure and bank risk (2010) Dutch National Bank Working Paper. Available at:\\ $http://www.dnb.nl/en/publications/dnb-publications/dnb-working-papers-series/dnb-working-papers/dnb233863.jsp$ (Accessed: 06/11/2013).

\bibitem{iman4}
van den End, W. A. \emph{Credit and liquidity risk of banks in stress conditions. Analyses from a macro perspective} (University Library Groningen, Groningen, 2011).

\bibitem{bargall}
Bargigli, L. \& Gallegati, M. Finding Communities in Credit Networks, \emph{Economics: The Open-Access, Open-Assessment E-Journal} {\bf 7}, 2013-17 (2013). Available at:\\ $http://dx.doi.org/10.5018/economics-ejournal.ja.2013-17$ (Accessed: 06/11/2013).

\bibitem{bargall2}
Bargigli, L. \& Gallegati, M. Random digraphs with given expected degree sequences: a model for economic networks, \emph{J. Econ. Behav. Organ.} {\bf 78}, 396-411 (2011).

\bibitem{simon}
Wells, S., Financial interlinkages in the United Kingdom's interbank market and the risk of contagion (2004) Bank of England Working Paper. Available at:\\ $http://www.bankofengland.co.uk/research/Documents/workingpapers/2004/WP230.pdf$ (Accessed: 06/11/2013).

\bibitem{myPRE1}
Squartini, T., Fagiolo, G. \& Garlaschelli, D. Randomizing world trade. I. A binary network analysis, \emph{Phys. Rev. E} {\bf 84}, 046117 (2011).

\bibitem{myPRE2}
Squartini, T., Fagiolo, G. \& Garlaschelli, D. Randomizing world trade. II. A weighted network analysis, \emph{Phys. Rev. E} {\bf 84}, 046118 (2011).

\bibitem{mynull}
Squartini, T., Fagiolo, G. \& Garlaschelli, D. Null models of economic networks: the case of the world trade web, \emph{J. Econ. Interact. Coord.} {\bf 8}, 75-107 (2012).

\bibitem{mattmars}
Mastromatteo, I., Zarinelli, E. \& Marsili, M. Reconstruction of financial networks for robust estimation of systemic risk, \emph{J. Stat. Mech} {\bf 03}, P03011 (2012).

\bibitem{simulations}
Upper, C. Simulation results to assess the danger of contagion in interbank markets, \emph{J. Financ. Stability} {\bf 7}, 111-125 (2011).

\bibitem{debtrank}
Battiston, S., Puliga, M., Kaushik, R., Tasca, P. \& Caldarelli, G. DebtRank: too central to fail? Financial networks, the FED and systemic risk, \emph{Sci. Rep.} {\bf 2} (2012).

\bibitem{eisenberg}
Eisenberg, L. \& Noe, T. H. Systemic risk in financial systems, \emph{Manag. Science} {\bf 47}, 236-249 (2001).

\bibitem{myfoodwebs}
Garlaschelli, D., Caldarelli, G. \& Pietronero, L. Universal scaling relations in food webs, \emph{Nature} {\bf 423}, 165-168 (2003).

\bibitem{foodwebmotifs}
Stouffer, D. B., Camacho, J., Jiang, W. \& Nunes Amaral, L. A. Evidence for the existence of a robust pattern of prey selection in food webs, \emph{Proceed. Royal Soc. B: Biol. Sci.} {\bf 274}, 1931-1940 (2007).

\bibitem{co-pierre}
Georg, C.-P., The effect of the interbank network structure on contagion and common shocks, \emph{J. Bank. Financ.} {\bf 7}, 2216-2228 (2013).

\bibitem{mymethod}
Squartini, T. \& Garlaschelli, D. Analytical maximum-likelihood method to detect patterns in real networks, \emph{New. J. Phys.} {\bf 13}, 083001 (2011).

\bibitem{newman_expo}
Park, J. \& Newman, M. E. J. The statistical mechanics of networks, \emph{Phys. Rev. E} {\bf 70}, 066117 (2004).

\bibitem{motifs}
Milo, R., Shen-Orr, S., Itzkovitz, S., Kashtan, N., Chklovskii, D. \& Alon, U. Network motifs: simple building blocks of complex networks, \emph{Science} {\bf 298}, 824-827 (2002).

\bibitem{motifs2}
Milo, R., Itzkovitz, S., Kashtan, N., Levitt, R., Shen-Orr, S., Ayzenshtat, I., Sheffer, M. \& Alon, U. Superfamilies of evolved and designed networks, \emph{Science} {\bf 303}, 1538-1542 (2004).

\bibitem{mymotifs}
Squartini, T. \& Garlaschelli, D. Triadic Motifs and Dyadic Self-Organization in the World Trade Network, \emph{Lec. Notes Comp. Sci.} {\bf 7166}, 24-35 (2012).

\bibitem{representative}
Kirman, A. P. Whom or what does the representative individual represent?, \emph{J. Econ. Perspect.} {\bf 6}, 117-136 (1992).

\bibitem{acharya}
Acharya, V. V. \& Bisin, A. Counterparty risk externality: centralized versus over-the-counter markets (2011) National Bureau of Economic Research Working Paper. Available at: $http://www.nber.org/papers/w{\it 17000}$ (Accessed: 06/11/2013).

\bibitem{Caruana}
Caruana, J., Interconnectedness and the importance of international data-sharing, Bank of International Settlements Speech. Paper presented at \emph{3rd Swiss National Bank - International Monetary Fund conference on the international monetary system}, Z\"urich (2012/05/08). Available at: $http://www.bis.org/speeches/sp120730.htm$ (Accessed: 06/11/2013).

\bibitem{Langfield}
Langfield, S., Liu, Z. \& Ota, T. Mapping the UK interbank system, Bank of England (2012) European Systemic Risk Board and UK Financial Services Authority Working Paper. Available at:\\ $http://catedrabde.uji.es/naim/images/presentations/langfield\_liu\_ota\_2012\_mapping\_the\_uk\_interbank\_system\_v110.pdf$ (Accessed: 06/11/2013).

\bibitem{Freixas}
Freixas, X., Parigi, B. \& Rochet, J. C. Systemic Risk, Interbank Relations and Liquidity Provision by the Central Bank, \emph{J. Money Credit Bank.} {\bf 3}, 611-638 (2000).

\bibitem{shannon}
Shannon, C. A Mathematical Theory of Communication, \emph{Bell System Tech. Jour.} {\bf 27}, 379-423, 623-656 (1948).

\bibitem{jaynes}
Jaynes, E. T. Information Theory and Statistical Mechanics, \emph{Phys. Rev.} {\bf 106}(4) (1957).

\bibitem{HL}
Holland, P. \& Leinhardt, S. \emph{Sociological Methodology}, Heise (ed.) (San Francisco, 1975).

\bibitem{WF}
Wasserman, S. \& Faust, K. \emph{Social Network Analysis} (Cambridge University Press, New York, 1994).

\bibitem{newman_expo}
Park, J. \& Newman, M. E. J. The statistical mechanics of networks, \emph{Phys. Rev. E} {\bf 70}, 066117 (2004).

\bibitem{mylikelihood}
Garlaschelli, D. \& Loffredo, M. I. Maximum likelihood: extracting unbiased information from complex networks, \emph{Phys. Rev. E} {\bf 78}, 015101(R) (2008).

\bibitem{myrec1}
Garlaschelli, D. \& Loffredo, M. I. Patterns of link reciprocity in directed networks, \emph{Phys. Rev. Lett.} {\bf 93}, 268701 (2004).

\bibitem{myrec2}
Garlaschelli, D. \& Loffredo, M. I. Multispecies grand-canonical models for networks with reciprocity, \emph{Phys. Rev. E} {\bf 73}, 015101(R) (2006).

\bibitem{newrec}
Newman, M. E. J., Forrest, S. \& Balthrop, J. Email networks and the spread of computer viruses, \emph{Phys. Rev. E} {\bf 66}, 035101(R) (2002).

\bibitem{model_selection}
Burnham, K. P. \& Anderson, D. R. \emph{Model Selection and Inference: A Practical Information-Theoretical Approach} (Springer-Verlag, New York, 1998).

\bibitem{mydistances}
Picciolo, F., Squartini, T., Ruzzenenti, F., Basosi, R. \& Garlaschelli, D. The role of distances in the World Trade Web, in \emph{Proceedings of the 2012 IEEE/ACIS 11th International Conference on Computer and Information Science}, 784-792 (IEEE, 2013).

\end{thebibliography}
\end{document}